\documentclass[aps,showpacs,nofootinbib,superscriptaddress]{revtex4-1}
\usepackage{graphicx}				
\usepackage{epsf}
\usepackage{mathrsfs}
\usepackage{epstopdf}
\usepackage{amssymb}
\usepackage{amsmath}
\usepackage{amsfonts}
\usepackage{amsthm}
\usepackage{pgfplots}
\usepackage{subfig}
\usepackage{epstopdf}
\usepackage{slashed}
\usepackage{wrapfig}
\usepackage{hyperref}
\hypersetup{
    colorlinks=true,
    linkcolor=blue,
    filecolor=magenta,      
    urlcolor=cyan,
    citecolor=blue,
}

\begin{document}

\title{Accuracy of the lepton-nucleus scattering models in the quasielastic region}
\date{\today} 
\begin{abstract}
We present a discussion of models of nuclear effects used to describe an inclusive electron-nucleus scattering in the quasielastic (QE) peak region, aiming to compare them and draw conclusion of their reliability when applied in neutrino-nucleus interactions. A basic motivation is to reduce systematic errors in neutrino oscillation experiments. We concentrate on the neutrino energy profile of the T2K experiment, which provides us a region of the greatest importance in terms of the highest contribution to the charge-current quasielastic (CCQE) cross section. We choose only electron-nucleus data that overlaps with this region. 
In order to clearify the analysis, we split the data sets into three groups and draw conclusion separately from each one of them.

We selected six models for this comparison: Benhar's spectral function with and without final state interaction (Benhar's SF+FSI), Valencia spectral function (Valencia SF), for higher energy transfer only with the hole spectral function; GiBUU and also the local and global Fermi gas models. The latter two are included as a benchmark to quantify the effect of various nuclear effects. All the six models are often used in neutrino scattering studies. A short theoretical description of each model is given. Although in the selected data sets the QE mechanism dominates, we also briefly discuss a possible impact of $2p2h$ and $\Delta$ contributions.

\end{abstract}
\author{Joanna E. Sobczyk}
\affiliation{Faculty of Physics and Astronomy, University of Wroclaw, Poland}
\maketitle

\section{Introduction}
The neutrino oscillation experiments - both currently running and those designed to work in the near future -  aim to measure the oscillation parameters with an increasing accuracy. The task is demanding as the measurement of the CP violation phase requires systematic errors to be at the level of $1-2\%$ \cite{Alvarez-Ruso:2017oui}. An important contribution to  systematic errors comes from models for neutrino-nucleus interactions used in Monte Carlo generators \cite{Ankowski:2016jdd}. This motivated many theoretical groups to work in this field, proposing formalisms that describe various channels of neutrino-nucleus scattering (quasielastic mechanism, $2p2h$, $\Delta$ production, etc.) taking into account nuclear effects. 

A natural question appears of how large the theoretical uncertainty in Monte Carlo simulations is and what is the theoretical limit of this uncertainty below which we cannot reach with the existing models. One cannot answer this question in a straightforward way, because no verification of the model predictions against the data can be done in the case of neutrinos. Not only they interact weakly, but also the neutrino beam is never monoenergetic which blurs the final analysis. However, there is a possibility to compare existing models of nuclear effects with the electron data which does not suffer from those two drawbacks. Obviously, for electrons the axial part of the current is missing and therefore there are only two structure functions in the hadron tensor, making the vertex of interaction less complicated than in the neutrino case. Nevertheless, the limit for theoretical uncertainty can  be only higher in the case of neutrino-nucleus interaction. 

The aim of this work is to estimate the uncertainty of theoretical models that describe electron scattering on nuclei. This depends on the electron energy (or more precisely on the kinematics of the process). We would like to concentrate our analysis on the energy range of the T2K experiment \cite{Abe:2011ks, Abe:2011sj, Abe:2013hdq}. For the neutrino beam used there, the main contribution to the neutrino-nucleus cross section comes from the charge-current quasielastic (CCQE) mechanism: 
\begin{eqnarray}
\begin{split}
&\nu_l + n \rightarrow l^- +p \\
&\bar{\nu_l} + p \rightarrow l^+ +n
\end{split}
\end{eqnarray}
where $l$ stands for lepton (electron, muon or tau), $p,n$ are proton and neutron.
We will focus on this reaction for muon neutrino. We also choose $^{12}$C as a target because it is the main target of the close detector in the T2K experiment \cite{Abe:2011ks}. Moreover, there are much more electron experimental data for $^{12}$C than for $^{16}$O, which is a main target in the T2K far detector.

For the comparison we have chosen models which are often used in the neutrino-nucleus interaction studies: Benhar's spectral function model (both with and without final state interactions) \cite{Benhar:1994hw, Benhar:2010nx, Petraki:2002nb, Ankowski:2014yfa}, Valencia spectral function \cite{Gil:1997bm, Nieves:2004wx, Nieves:2017lij, FernandezdeCordoba:1991wf} and GiBUU \cite{Buss:2011mx, Gallmeister:2016dnq}. We also include predictions of global and local Fermi gas models, as they are a ''starting point'' for the construction of the abovementioned models. In this way we may observe what is the role of the nuclear effects for various kinematical regions.

The paper is organized as follows. In Sec. \ref{sec:selection} we describe the selection of the electron data used for the comparison. We introduce a criterium for importance of data sets from the point of view of  the T2K beam \cite{Abe:2012av}, dividing them accordingly into three groups. This will allow us to perform a more systematic analysis of nuclear effects.  In Sec. \ref{sec:models} we present the main features of theoretical models that will be discussed in this paper. In Sec. \ref{sec:results} we proceed with an analysis of the models separately for each energy transfer region. 
Sec. \ref{sec:discussion} and \ref{sec:conlusions} contain a discussion of the results  and conclusions.


\section{Electron data selection}\label{sec:selection}
\subsection{Regions}\label{sec:regions}
\begin{figure}[h]
\centering
\includegraphics[scale=0.4]{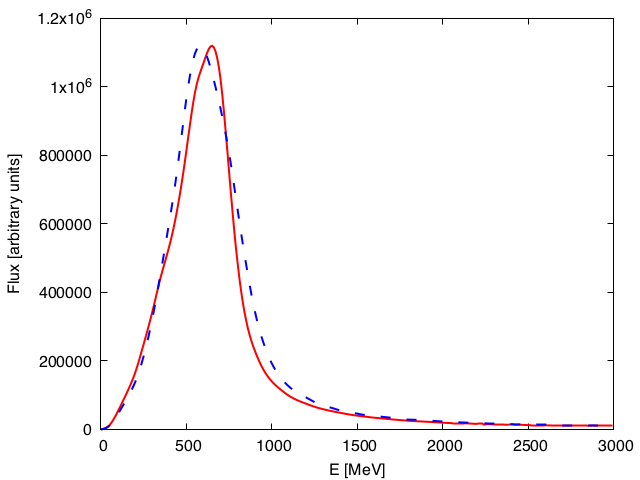}
\caption{Muon neutrino flux in ND280 detector of T2K experiment, taken from \cite{Abe:2012av}. The dashed line shows the flux, the solid curve shows flux multiplied by the oscillation probability $\mathcal {P}_{\nu_{\mu} \rightarrow \nu_{e}}$ \cite{Abe:2014ugx}. Both are normalized to the same area.}
\label{fig:flux}
\end{figure}

From a wide range of kinematics of electron-carbon scattering experiment data, we would like to choose only those data sets that are the most important for the T2K experiment 
In order to do this, we calculate an estimated value of flux-averaged differential cross section $\frac{d\sigma}{dq d\omega}^{CCQE}$ where $(\omega, |\vec{q}|)$ is energy-momentum transfer. The differential cross section has to be averaged over the muon neutrino flux used in the T2K experiment $\mathcal{F}(E)$ shown on Fig. \ref{fig:flux}. The cross section averaged over the flux is given by:
\begin{equation}\label{eq:t2k_a}
\frac{d\sigma}{dq d\omega}_{T2K} =\frac{1}{\mathcal{F}} \int d E \frac{d\sigma}{dq d\omega}^{CCQE} \mathcal{F}(E), \ \ \ \ \ \ \mathcal{F} = \int dE \mathcal{F}(E)
\end{equation}
 
Afterwards we look at the distribution of oscillated neutrinos investigated in both appearance and disappearance measurements, taking into account the probability of oscillation $\mathcal {P}_{\nu_{\mu} \rightarrow \nu_{e}}$ with parameters given in \cite{Abe:2014ugx}. It shifts slightly the peak and makes it narrower (see Fig. \ref{fig:flux}). The cross section has the form:
\begin{equation}\label{eq:t2k}
\frac{d\sigma}{dq d\omega}_{T2K}^{\text{osc}} =\frac{1}{\mathcal{W}} \int d E \frac{d\sigma}{dq d\omega}^{CCQE} \mathcal{F}(E) \mathcal {P}_{\nu_{\mu}\rightarrow \nu_{e}} (E), \ \ \ \ \ \ \mathcal{W}= \int d E  \mathcal{F}(E) \mathcal {P}_{\nu_{\mu}\rightarrow \nu_{e}} (E)
\end{equation}
The results of the analysis done with Eq. \ref{eq:t2k_a} and \ref{eq:t2k} are very similar because the spectra for muon neutrinos and oscillated neutrinos are very much alike (see Fig. \ref{fig:flux}). In the further analysis we will choose the second option, Eq. \ref{eq:t2k}.

In order to calculate $\frac{d\sigma}{dq d\omega}^{CCQE}$ we use Valencia SF model (with a free particle spectral function, see Subsec. \ref{sec:valencia}) because of its fast numerical performance. 

\begin{figure}[h]
\centering
\includegraphics[scale=0.45]{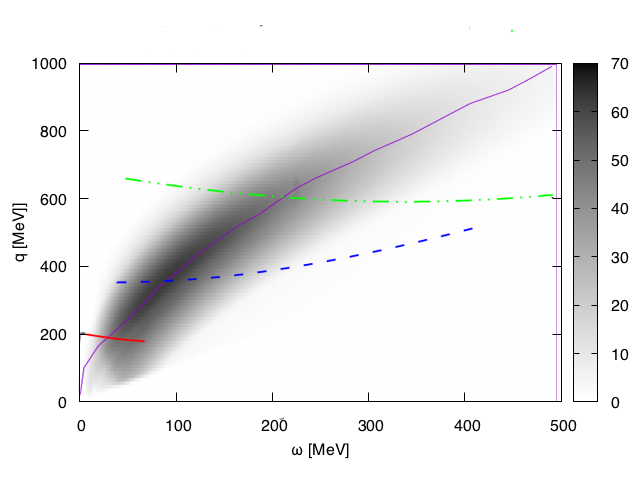}
\caption{$\frac{d\sigma}{dq dq_0}_{T2K}^{\text{osc}}$ [$ 10^{-38}\frac{cm^2}{GeV^2}$] for the T2K flux (using prediction of Valencia spectral function), see Eq. \ref{eq:t2k}. With the solid line the position of the CCQE peak is marked. Three typical  electron data sets are also shown (experimental points are connected in line for better legibility): solid line E= 200 MeV $\theta=60^\circ$; dashed line E= 1500 MeV $\theta=13.54^\circ$, dotted-dashed line E= 680 MeV $\theta=60^\circ$. Each of them cross the solid line (''CCQE line'') in a different point. The value of  $\frac{d\sigma}{dq d\omega}_{T2K}^{\text{osc}}$ in this crossing point is taken as a measure of importance of the data set. 
}
\label{fig:dqdw}
\end{figure}

On the resulting plot (see Fig. \ref{fig:dqdw}) we mark the position of the QE peak, defined as a maximum value of $\frac{d\sigma}{dq d\omega}^{CCQE}$ for a given $\omega$ (note that its position is  very close to $\vec{q}^2=\sqrt{\omega^2+2M\omega}$). 
There is some model dependence in the shape shown of Fig.  \ref{fig:dqdw}: the usage of different models might slightly change it (e.g. shift the position of the QE peak, spread it, etc.) however it would not influence our conclusions since we use it only to sieve the electron data.

There are in total 66 data sets of electron-carbon scattering data gathered in \cite{Benhar:2006er}. They cover a wide range of incoming electron energy and scattering angle which can be translated into energy-momentum transfer. With this change of variables, we can plot them all on the $(\omega, |\vec{q}|)$ plane, just as shown on Fig. \ref{fig:dqdw} for three example data sets. The usage of $(\omega, |\vec{q}|)$ variables is more natural because thry are arguments of nuclear response functions, which are analogical for neutrino and electron scattering processes (the difference lies in the interaction vertex which makes a value of weak response functions larger). Therefore, on Fig. \ref{fig:dqdw} we show how electron sets coincide with the CCQE region for the T2K flux. 

\begin{figure}[h]
\includegraphics[scale=0.35]{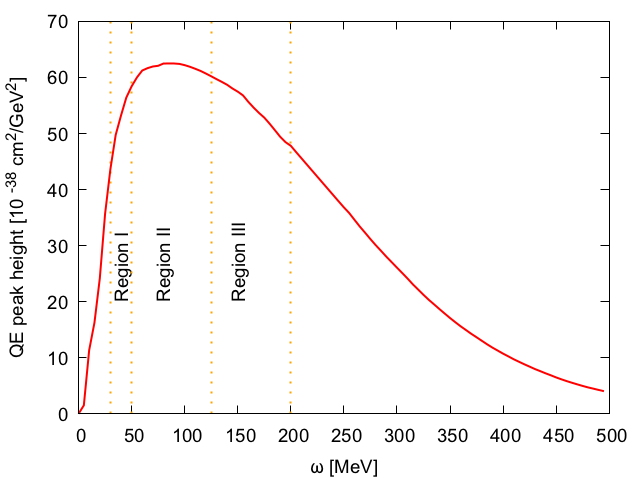}
\caption{A profile of the CCQE peak (the profile along the line marked on Fig. \ref{fig:dqdw} projected on the $\omega$ axis).}
\label{fig:dqdw_profile}
\end{figure}

The next problem is to define a criterium which data sets are most important for neutrino studies. We propose to quantify the significance of data using the following procedure (points 1-3 are explained above):
\begin{enumerate}
\item We calculate $\frac{d\sigma}{dq d\omega}_{T2K}^{\text{osc}}$ for the range of $\omega\in (0,500)$ MeV, $|\vec{q}|\in (0,1000)$ MeV/c.
\item The QE peak is defined as the maximal value of $\frac{d\sigma}{dq d\omega}_{T2K}^{\text{osc}}$ for a given $\omega$ (it forms a "CCQE line" marked as solid line on Fig. \ref{fig:dqdw} )
\item We plot electron data sets on the ($\omega$, $|\vec{q}|$) plane.
\item We decide to choose only those data sets whose CCQE value (the value of $\frac{d\sigma}{dq d\omega}_{T2K}^{\text{osc}}$ at the crossing point with the ''CCQE line'') is at least $67\%$ of the maximal value in the ''CCQE line'', see Fig. \ref{fig:dqdw_profile}.
\end{enumerate}

This procedure largely reduces all the available electron scattering data, basically cutting off those with high-energy transfer and very low-energy transfer in the QE peak. There are in total 32 data sets fulfilling our  condition. To introduce further order among them, we divide them into three groups according to their importance, meaning again the value of the CCQE peak that corresponds to their position. On Fig.  \ref{fig:dqdw_profile} we plot the dependence of the CCQE peak height on the energy transfer $\omega$ (using Valencia spectral function model) and mark three regions. In the energy range $30-200$ MeV the height varies from 43.7 to 62.5$\times10^{-38}\frac{cm^2}{GeV^2}$ . The maximum value is reached for $\sim 85$ MeV. Three regions are:
\begin{itemize}
\item \textbf{Region I (energy transfer 30-50 MeV)}. The momentum transfer in this region is $|\vec{q}|<300$ MeV, which means that actually we are beyond the limit of usability of the Impulse Approximation and giant resonances are visible in the data (see Sec. \ref{sec:models}). This is also the smallest sample with only 5 data sets.
\item \textbf{Region II (energy transfer 50-125 MeV)}. This would be our main region of interest, it covers data with at least $95\%$ of the maximal height. There are 18 data sets in this region.
\item \textbf{Region III (energy transfer 125-200 MeV)}. The importance of this region is similar to Region I, however we separate them for the sake of clearer analysis (as theoretical models behave differently in Region I and III). There are 9 data sets in Region III.
\end{itemize}

One might think of different procedures to choose the data. E. g. instead of looking at the highest point (''CCQE line''), one could mark the area of the highest cross section and look for the integrated electron cross section in the overlap. This is going to be a subject of the subsequent study.
\subsection{Data sets}\label{sec:sets}
Selected 32 data sets are presented in Tables \ref{table:electrons1} (Region II) and \ref{table:electrons2} (Region I and III). Each data set is labeled with the energy of incoming electron and scattering angle (columns 1 and 2). 

For each data set  we find a point $(\omega, |\vec{q}|)$ for which $\frac{d\sigma}{dq d\omega}_{T2K}^{\text{osc}}$ is maximal.  Therefore, we denote point coordinates as ($\omega^{QE}$,  $|\vec{q}|^{QE}$) and show the values in columns 3, 4. Then we calculate $\frac{d\sigma}{dq d\omega}_{T2K}^{\text{osc}} (\omega^{QE}$,  $|\vec{q}|^{QE})$ (column 5). Finally, we calculate the ratio of this value and  the maximal value (among all the sets it corresponds to kinematics E=1650 MeV, $\theta=13.54^{\circ}$). This serves us as a measure of importance of the set, and is shown in column 6.

For the future discussion we estimate contributions from longitudinal and transverse response in the experimental data.  For electron-nucleus scattering, we can present cross section (see Eq. \ref{eq:1}) as a sum of two contributions from either transverse or longitudinally polarized virtual photon. It can be written by means of  $R_T(q)$, $R_L(q)$ response functions:
\begin{equation}
\frac{d^2\sigma}{d\Omega d \omega} = \bigg( \frac{d \sigma}{d\Omega}\bigg)_{Mott} \bigg[  \frac{Q^4}{\vec{q}^4} R_L(q) + \big(\frac{1}{2}\frac{Q^2}{\vec{q}^2}+\tan^2\frac{\theta}{2}\big)R_T(q)  \bigg]  =  \bigg( \frac{d \sigma}{d\Omega}\bigg)_{Mott} \bigg[  \sigma_L + \sigma_T  \bigg]
\end{equation}
where $Q^2 = -q^2$. It is known that non-QE processes, $2p2h$ and $\Delta$ production, give raise to $R_T$ part. Therefore, for $\sigma_L\gg\sigma_T$ we might expect those mechanisms to give small contribution in the QE region. We will come back to this point in Sec. \ref{sec:discussion}. In column 7 we show $\sigma_T/\sigma_L$. Its value calculated at the peak tells us if one should expect large contribution from $2p2h$ and $\Delta$ mechanisms.

 \begin{table*}[h]
\centering
\begin{tabular}{|c c | c c | c |c | c |}
  \hline
Energy [MeV] & Angle [Degrees] & $\omega^{QE}$ [MeV]  & $|\vec{q}|^{QE}$ [MeV]  & $\frac{d\sigma}{dq d\omega}_{T2K}^{\text{osc}}(\omega^{QE}, |\vec{q}|^{QE})$ & $ \frac{\text{height}}{\text{max\ height}} [\%]$ & $\frac{\sigma_T}{\sigma_L}(\omega^{QE}, |\vec{q}|^{QE})$\\
& & & & [$10^{-38}$ cm$^2$/GeV$^2$] & & \\
\hline
280	&	60	&	53	&	258	&	60.1	&	96.1	&	0.4	\\
1300	&	11.95	&	60	&	271	&	61.2	&	98.0	&	0.2	\\
480	&	36	&	62	&	284	&	61.5	&	98.3	&	0.3	\\
320	&	60	&	63	&	294	&	61.3	&	98.1	&	0.5	\\
1500	&	11.95	&	70	&	313	&	61.7	&	98.8	&	0.3	\\
1300	&	13.54	&	70	&	306	&	61.9	&	99.1	&	0.3	\\
560	&	36	&	72	&	331	&	62.2	&	99.4	&	0.4	\\
361	&	60	&	73	&	331	&	62.2	&	99.5	&	0.6	\\
1650	&	11.95	&	80	&	345	&	62.1	&	99.4	&	0.4	\\
1500	&	13.54	&	80	&	353	&	62.4	&	99.9	&	0.4	\\
401	&	60	&	88	&	365	&	62.3	&	99.7	&	0.7	\\
620	&	36	&	92	&	365	&	62.3	&	99.7	&	0.5	\\
440	&	60	&	98	&	400	&	61.7	&	98.7	&	0.8	\\
1650	&	13.54	&	100	&	390	&	62.0	&	99.2	&	0.4	\\
680	&	36	&	102	&	401	&	61.5	&	98.5	&	0.6	\\
480	&	60	&	112	&	435	&	60.5	&	96.9	&	0.9	\\
730	&	37.1	&	115	&	442	&	60.2	&	96.3	&	0.7	\\
500	&	60	&	125	&	451	&	59.9	&	95.8	&	1.0	\\
   \hline
 \hline
\end{tabular}
\caption{ The data sets from Region II sorted by $\omega^{QE}$. Columns 3 and 4: the point ($\omega^{QE}$,  $|\vec{q}|^{QE}$) in which $\frac{d\sigma}{dq d\omega}_{T2K}^{\text{osc}}$ is the highest for the given data set. Column 5: the highest value of $\frac{d\sigma}{dq d\omega}_{T2K}^{\text{osc}}$. Column 6: the ratio of this value ('height'') and  the maximal value, ''max height'' (among all the sets it corresponds to kinematics E=1650 MeV, $\theta=13.54^{\circ}$). The numbers given in the last column where obtained in the local Fermi gas model. For a detailed explanation of the table see Subsec. \ref{sec:sets}.}
\label{table:electrons1}
\end{table*}

 \begin{table*}[h]
\centering
\begin{tabular}{|c c | c c | c |c | c |}
\hline
Energy [MeV] & Angle [Degrees] & $\omega^{QE}$ [MeV]  & $|\vec{q}|^{QE}$ [MeV] & $\frac{d\sigma}{dq d\omega}_{T2K}^{\text{osc}} (\omega^{QE}, |\vec{q}|^{QE})$&  $\frac{\text{height}}{\text{max\ height}} [\%]$ & $\frac{\sigma_T}{\sigma_L}(\omega^{QE}, |\vec{q}|^{QE})$\\
& & & & [$10^{-38}$ cm$^2$/GeV$^2$] & & \\
\hline
200	&	60	&	43	&	182	&	50.3	&	80.5	&	0.2	\\
320	&	36	&	43	&	189	&	51.9	&	83.0	&	0.2	\\
240	&	36	&	48	&	141	&	44.7	&	71.5	&	0.1	\\
240	&	60	&	48	&	220	&	56.8	&	90.9	&	0.3	\\
400	&	36	&	52	&	236	&	58.3	&	93.3	&	0.2	\\
\hline
519	&	60	&	131	&	468	&	59.0	&	94.3	&	1.0	\\
560	&	60	&	142	&	504	&	57.0	&	91.2	&	1.2	\\
2020	&	15.022	&	150	&	530	&	54.2	&	86.7	&	0.7	\\
2000	&	15	&	150	&	524	&	55.0	&	88.0	&	0.7	\\
1930	&	16	&	164	&	539	&	53.2	&	85.1	&	0.8	\\
620	&	60	&	168	&	555	&	51.6	&	82.6	&	1.3	\\
2130	&	16	&	179	&	595	&	47.2	&	75.5	&	0.9	\\
1930	&	18	&	182	&	603	&	46.0	&	73.7	&	0.9	\\
961	&	37.5	&	183	&	585	&	48.7	&	78.0	&	1.0	\\
680	&	60	&	188	&	608	&	46.0	&	73.7	&	1.5	\\
   \hline
 \hline
\end{tabular}
\caption{As in Table \ref{table:electrons1}. The data samples from Region I and III sorted by $\omega^{QE}$. The first 5 sets belong to Region I, remaining to Region III. }
\label{table:electrons2}
\end{table*}

\section{Theoretical models}\label{sec:models}
Models presented in this section describe lepton-nucleus interaction where lepton interacts with only one nucleon out of a nucleus (Impulse Approximation). This approximation is valid only for momentum transfer high enough to penetrate nucleus structure at $\sim 300$ MeV/c. Below this threshold it becomes less founded, thus we should not expect models presented below to give a good description of the scattering. In the low momentum transfer region interaction does not take place on a single nucleon. It is rather a collective excitation in which many nucleons take part. This kind of phenomena is described by CRPA or RPA techniques \cite{Gil:1997bm, Martini:2016eec, Kolbe:2003ys}. We will come back to this problem again while analyzing the models' predictions in Sec. \ref{sec:results}. Also we should note that two body mechanism (so called $2p2h$ or MEC) - and to a less extent $\Delta$ excitation - can give non-negligible contributions in the QE region. It is when transverse contribution is large. We address this problem in Sec. \ref{sec:2p2h}.

In order to properly describe nucleons, which are not free particles in the nucleus, a formalism of spectral functions is used. The spectral function $P^{h,p}(E, p)$ describes either the hole state (a nucleon of energy below Fermi sea) or the particle state (above Fermi sea). For the hole state it can be spelled out as a probability density of removing a nucleon of momentum p leaving nucleus with the removal energy E. In a free case it is simply $\delta(E-\sqrt{M^2+p^2})$, where $M$ is nucleon mass, but nuclear corrections greatly change it. Below we will present each model in terms of spectral functions. They enter the formula for the cross section in the standard way:
\begin{equation}\label{eq:1}
\frac{d^2\sigma}{d\Omega d \omega} =\bigg( \frac{\alpha}{Q^2}\bigg)^2 \frac{|\vec{k}|}{|\vec{k}'|} L_{\mu\nu} W^{\mu\nu}
\end{equation}
where $\vec{k}$, $\vec{k}'$ are incoming and outcoming lepton momenta, $L_{\mu\nu}$, $W^{\mu\nu}$ are respectively lepton and hadron tensor. Latter can be written in terms of the spectral functions:
\begin{eqnarray}
\begin{split}
W^{\mu\nu}(q) = & \int d^3r \rho(r) \int \frac{d^3p}{(2\pi)^3} \frac{3}{4\pi k_F(\rho)^3}  \frac{M}{E_p}\frac{M}{E_{p+q}} \int d E P^{h}(E,\vec{p}; \rho) P^{p}(\omega-E,\vec{p}+\vec{q}; \rho)A^{\mu\nu}(p,q)
\end{split}
\label{eq:hadron_tens}
\end{eqnarray}
$ A^{\mu\nu}(p,q)$ is a tensor describing interaction with a single nucleon. 

\subsection{Fermi gas}
Fermi gas is a naive model, as it only introduces constant binding energy for nucleon-nucleon interaction in the nucleus and takes into account statistical correlations in the Fermi system. However, it is still useful to know in what energy range is works well and where it breaks down. As one can suspect, especially for low energy transfer where nuclear effects are pronounced, this model becomes helpless, largely overestimating the QE peak and displacing it. Its main advantage is simplicity which translates into a very fast numerical computation. Spectral functions for the Fermi gas have a very simple form (respectively for the hole and particle state)\footnote{For simplicity let us assume a symmetric nucleus with the same spectral function for protons and neutrons.}:
\begin{eqnarray}
\begin{split}
&P_{FG}^{h} (E, p; \rho)= \theta\big(k_F(\rho)-p\big) \delta \big(E+\sqrt{M^2+p^2}-M - B\big)\\
&P_{FG}^{p} (E, p; \rho)=\theta\big(p-k_F(\rho)\big) \delta \big(E-\sqrt{M^2+p^2}+M\big)
\label{eq:lfg}
\end{split}
\end{eqnarray}
where $B$ is the binding energy, usually fitted to experimental data.
Thus the hadron tensor takes a form:
\begin{eqnarray}
\begin{split}
W^{\mu\nu}(q) =  \int d^3r \rho(r) \int \frac{d^3p}{(2\pi)^3} \frac{3}{4\pi k_F(\rho)^3}  \frac{M}{E_p}\frac{M}{E_{p+q}} \theta\big(k_F(\rho)-|\vec{p}|\big) \theta\big(|\vec{p}+\vec{q}|-k_F(\rho)\big) \delta\big( E_{p+q}-E_p-\omega+B \big) A^{\mu\nu}(p,q)
\end{split}
\label{eq:fg}
\end{eqnarray}
where $k_F(\rho)$ is the Fermi momentum (which is $\rho$ density dependant in the case of local Fermi gas) and $E_p$, $E_{p+q}$ are kinetic energies.

The integration over volume has to be done only for the local Fermi gas.
In the case of the global Fermi gas (GFG) we set Fermi momentum to 221 MeV/c. The binding energy is given by 25 MeV for both the GFG and LFG.

\subsection{Benhar's spectral function (SF)}
In the Plane Wave Impulse Approximation (PWIA), the lepton-nucleus scattering is approximated by the lepton-nucleon interaction where a struck nucleon can be treated as a free particle (it is described by the spectral function $P_{FG}^{p}(p,E)$ from Eq. \ref{eq:lfg}). This assumption is applicable in many cases when the momentum transfer is high enough so that the residual interaction between the nucleon and the medium can be neglected (for more explanation see e.g. \cite{Benhar:2013dq}). The primary interacting nucleon (a hole state) is submerged in the nuclear medium, which changes its vacuum properties (e.g. introducing width and changing dispersion relations). Those effects are described in terms of a spectral function $P_{Benhar}(E,p)$ \cite{Benhar:1994hw}, where $\vec{p}$ is the nucleon momentum and $E$ is the removal energy. The hadron tensor is given by:
\begin{equation}
W^{\mu\nu}(q) = \int d^3p \int dE  \frac{M}{E_p}  \frac{M}{E_{p+q}}  P_{Benhar}(E, |\vec{p}|) P^{p}_{FG}(\omega-E+M-t_{A-1}, |\vec{p}+\vec{q}|) A^{\mu\nu}(p,q)
\label{eq:benharsf}
\end{equation}

The spectral function in Benhar's model consists of two parts. The main part is obtained within a shell-model, while the resting part is obtained by microscopic calculations carried out in the NMBT (Nuclear Many Body Theory) emplying Local Density Approximation. The calculation is nonrelativistic, which makes it apropriate only for a hole state (which has momentum lower than Fermi momentum $\approx 220$ MeV/c). The formalism describing the FSI (final state interaction) will be presented in the following subsection. For now only statistical correlations for the particle state will be used, which lead to Pauli blocking (here the momentum threshold is estimated to 211 MeV, an average Fermi momentum in local Fermi gas model).

One also should take into account the fact that the incoming and outcoming electron experience Coulomb field created by the nucleus. It distorts its wavefunction and thus influences the resulting cross section. There are several ways to include Coulomb corrections (for a short review see \cite{Benhar:2006wy}). For both Benhar's SF and SF with FSI (Sec. \ref{sec:fsi}) we shall use EMA' (improved effective momentum approximation) \cite{Aste:2004yz}. This effects however is negligible for light nuclei ($^{12}$C), changing the energy of an incoming electron by $\sim7$ MeV. The experimental data presented for the analysis contains the Coulomb effect.

\subsection{Benhar's SF with FSI}\label{sec:fsi}
In order to go beyond PWIA, an interaction between struck nucleon (a particle state) and the nuclear medium has to be modeled - not only for Pauli blocking but also dynamical correlations. There are several papers devoted to this formalism \cite{Benhar:1991af, Petraki:2002nb, Benhar:2013dq, Ankowski:2014yfa}. We will follow here a procedure given in \cite{Ankowski:2014yfa}, using Pauli blocking described in Eq. 10 of this reference. The FSI is treated in terms of an optical potential $U=U_W+i U_V$, which describes both a non-zero width of a nucleon in a nuclear medium ($U_V$ - the imaginary part of the potential) and a modification to its energy-momentum relation ($U_W$ - the real part of the potential, taken from the fit of \cite{Cooper:1993nx}). After some simplifications, the optical potential can be employed by means of a folding function $f(\omega)$ \cite{Benhar:2013dq} convoluted with the cross section from Eq. \ref{eq:benharsf}.
The strength of the FSI effect also depends on the nuclear transparency $T_A$ (which grows for the decreasing nucleon energy).

\begin{equation}
\frac{d\sigma^{FSI}}{d\omega d\Omega} = \int d\omega' f(\omega-\omega'-U_V)\frac{d\sigma^{SF}}{d\omega' d\Omega}
\end{equation}

\begin{equation}
f(\omega) = \delta(\omega)\sqrt{T_A} + (1-\sqrt{T_A}) \frac{1}{\pi}\frac{U_W}{U_W^2+\omega^2}
\end{equation}

Note that this formalism can be used in relativistic kinematics, which makes it suitable for processes at moderate and large energy transfer. On the other hand, for momentum transfer of the order of 1 GeV/c, the FSI becomes almost negligible.

\subsection{Valencia spectral function}\label{sec:valencia}
Spectral functions are also one of the ingredients of the formalism developed by Valencia group and used by Nieves et al. for neutrino interactions \cite{Nieves:2004wx,Nieves:2017lij}. In this approach the key approximation is the Local Density Approximation (LDA). One uses calculations done for nuclear medium of a constant density, integrating it over the nucleus density profile. This is a simpler approach then Benhar's model for the hole spectral function, but this simplicity also has an advantage: it can be employed for a variety of nuclei.

Spectral function $P_{Valencia}(E,p; \rho)$ of this model is obtained from a semi-phenomenological computation of a nucleon self-energy $\Sigma(E,p,\rho)$ in the nuclear medium developed in \cite{FernandezdeCordoba:1991wf}. The imaginary part of the self-energy was shown to give similar results as those obtained by microscopic calculations. The real part is calculated from dispersion relations, up to unknown constant term. This term, however, is not necessary in the computation where both hole and particle state are ''dressed'', and this is the case of the QE mechanism. Both the hole and the particle state are described by spectral functions, which enter the formula for the hadron tensor in the standard way from Eq. \ref{eq:hadron_tens}:
\begin{equation}
W^{\mu\nu}(q) = \int_0^{\infty} dr r^2 \frac{\Theta(\omega)}{4\pi^2}\int d^3p\int d E P_{Valencia}^{h}(E,\vec{p}\,; \rho) P_{Valencia}^{p}(E+\omega,\vec{p}+\vec{q}\,;\rho)A^{\mu\nu}(p,q)\bigg|_{p^0=E_p}
\end{equation}
and spectral functions are given by:
\begin{eqnarray}
\begin{split}
P^{h}_{Valencia}(E,\vec{p}\,; \rho) = - \frac{1}{\pi} \frac{\text{Im}\Sigma(E,\vec{p}\,)}{\big(E - M-\vec{p}^{\,2}/2M - \text{Re}\Sigma(E,\vec{p}\,) \big)^2 + \text{Im}\Sigma(E,\vec{p}\,)^2}  \theta\big(\mu-E \big) \\
P^{p}_{Valencia}(E,\vec{p}\,; \rho) = \frac{1}{\pi} \frac{\text{Im}\Sigma(E,\vec{p}\,)}{\big(E - M-\vec{p}^{\,2}/2M - \text{Re}\Sigma(E,\vec{p}\,) \big)^2 + \text{Im}\Sigma(\omega,\vec{p}\,)^2} \theta\big(E-\mu \big) \\
\end{split}
\end{eqnarray}
where $\mu$ is the chemical potential:
\begin{equation}
\mu(k_F) = M + \frac{k_F^2}{2M} + \text{Re}\Sigma (\mu(k_F) ,k_F) \label{eq:defchm}
\end{equation}

The main drawback of this approach is the nonrelativisic calculation of self-energies. This induces an error in the integrated cross section of approximately $7-10\%$ for the incoming lepton of energy 500 MeV (see \cite{Nieves:2017lij}). To show how much it affects the shape of differential cross sections, for results in Region I and II and partially III, we plot the LFG using both relativistic and non-relativistic kinematics. The effect is already getting sizeable for Region II of our analysis. 

In order to overcome this problem, there is a possibility to use only the hole spectral function, in the same manner as Benhar does. The details of how this can be done are explained in \cite{FernandezdeCordoba:1995pt} and recently used in \cite{Nieves:2017lij}. Dressing the particle state - or in other words including FSI effects - induces smaller changes for high energy transfers. Thus for Region III we will use the approximated version of the full model, e.i. $P^p_{FG}$ for the particle spectral function.

In the neutrino community Nieves et al. formalism is well-known not only because of spectral functions, but also thanks to incorporating RPA effects. However, inclusion of both nuclear effects introduces an additional uncertainty because RPA parameters were fitted for noninteracting nuclear system (for further information see \cite{Nieves:2017lij}). Moreover, RPA effects are parametrized by a spin-isospin interaction which is different in the case of electromagnetic, charge and neutral current interaction. Therefore it is difficult to conclude about RPA in neutrino induced processes basing on the electron scattering results. Here we will just show results for spectral functions. As was shown in previous papers, for integrated cross section inclusion of both RPA and SF does not much change the overall result in comparison with using SF alone.

Coulomb distortion is introduced in Valencia formalism in a way described in \cite{Singh:1993rg, Nieves:2004wx}. Even though it is a slightly different formalism than EMA', it gives substantially equal result (especially for light nuclei like carbon).

\subsection{GiBUU}
GiBUU (Giessen-Boltzmann-Uehling-Uhlenbeck) is an elaborated framework that provides a description of a great variety of nuclear processes \cite{Buss:2011mx}, among them lepton-nucleus interaction \cite{Leitner:2008ue}. It is based on the quantum-kinetic transport theory, allowing to explore both inclusive and exclusive processes. In this approach the key approximation is the Local Density Approximation (LDA) in which one uses calculation done for nuclear medium of a constant density, integrating it over the nucleus density profile. For the description of the nucleus ground state, nucleons' momenta are distributed according to the local Fermi gas model and submerged into momentum and position dependent potential $U(\vec{p}, \vec{r})$. It was obtained from an energy-density functional that reproduces the saturation for nuclear matter and was firstly used for the description of heavy-ion reactions:
\begin{equation}
U(\vec{p}, \vec{r}) = A \frac{\rho(\vec{r})}{\rho_0} + B \bigg( \frac{\rho(\vec{r})}{\rho_0} \bigg)^{\tau} + \frac{2 C}{\rho_0} g \int \frac{d^3p'}{(2\pi)^3}\frac{ \Theta(k_F-|\vec{p}|)}{1+(\frac{\vec{p}-\vec{p}'}{\Lambda})^2}
\end{equation}
where $\rho_0 = 0.168$ fm$^{-3}$ and $g = 4$.
GiBUU offers few possible parametrizations. The standard one uses $A = -29.3$ MeV, $B=57.2$ MeV, $C=-63.6$ MeV, $\tau = 1.76$ and $\Lambda = 2.13$ fm$^{-1}$.
This translates into an effective mass of the nucleons bounded in a nucleus:
\begin{equation}
M^*(\vec{p}, \vec{r}) = M+U(\vec{p}, \vec{r})
\end{equation}
or equivalently can be expressed in terms of spectral functions for nuclear matter:
\begin{eqnarray}
\begin{split}
&P_{GiBUU}^{h}(E,|\vec{p}|; \rho) =\theta\big(k_F(\rho)-p\big) \delta \big(E-M^*+\sqrt{M^{*2}+p^2}\big)\\
&P_{GiBUU}^{p}(E,|\vec{p}|; \rho) = \theta\big(p-k_F(\rho)\big) \delta \big(\sqrt{M^{*2}+p^2}-M^*-E\big)
\end{split}
\label{eq:gibuu}
\end{eqnarray}
In the above equations E is the removal energy for the hole spectral function (here we follow the notation of \cite{Gallmeister:2016dnq}) and for the particle spectral function E is the nucleon's kinetic energy. Even though the difference in the notation of $P_{GiBUU}^{p}$ and $P_{GiBUU}^{h}$ may be at first sight misleading, it was chosen in this way to correspond to notation of Eq. 8 from \cite{Ankowski:2014yfa}. 
In order to obtain an analogical form of the hole spectral function as in the model of Benhar \cite{Benhar:1994hw}, one has to perform integration over nucleus volume:
\begin{eqnarray}
\begin{split}
&\mathcal{P}_{GiBUU}^{h}(E,|\vec{p}|) = \int d^3 r P_{GiBUU}^{h}(E,|\vec{p}|; \rho)
\end{split}
\end{eqnarray}
$r$-dependence of the argument of the $\delta$ function in Eq. \ref{eq:gibuu} changes the dispersion relation in comparison to the Fermi gas model. After integration over volume in $\mathcal{P}_{GiBUU}^{h}$, it has not any more the form $E(p)=\sqrt{M^2+\vec{p}^2}-M$, but becomes smeared out.

The hadron tensor reads:
\begin{equation}
W^{\mu\nu}(q) = \int d^3r \int \frac{d^3p}{(2\pi)^3} \int d E  \frac{M}{E_p}  \frac{M}{E_{p+q}} P_{GiBUU}^{h}(E,\vec{p}\,; \rho) P_{GiBUU}^{p}(\omega-E,\vec{p}+\vec{q}\,;\rho)A^{\mu\nu}(p,q)
\end{equation}

In principle GiBUU uses spectral functions that take into account collisional broadening - the fact that nucleons interact with nuclear medium \cite{Lehr:2001qy}. This effect, however, is turned-off in the standard GiBUU simulation as it yields similar results for inclusive cross-sections. Not only the collisional spectral function has to be included for the hole state, but also for the final state interactions in order to involve off-shell nucleons. The latter effect requires more time-consuming computation (and it has to be taken into account when studying exclusive processes). The particle state thus ''feels'' the same mean-field potential as the hole state.

\section{Analysis}\label{sec:results}

In the following section we describe the results predicted by each model. In order to make plots legible, we do not include results of Benhar's SF without FSI. However, we mention its behaviour in the description. Also, its effect will be shown on figures included in Sec. \ref{sec:discussion}. 

\subsection{I Region (energy transfer 30-50 MeV)}\label{sec:30-50}
\begin{figure*}[h]
\centering
\subfloat[]{
\includegraphics[scale=0.35]{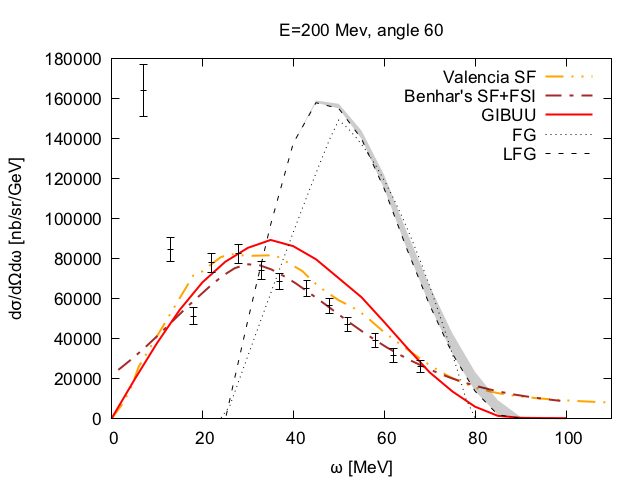}
\label{fig:30-50_aa}
}
\subfloat[]{
\includegraphics[scale=0.35]{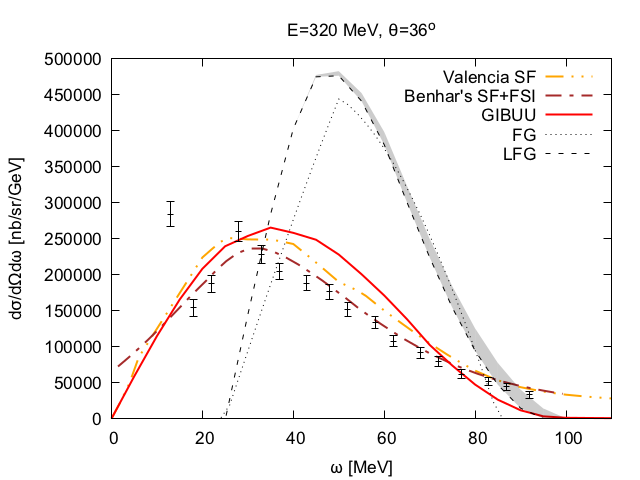}
\label{fig:30-50_ab}
}
\hspace{0mm}
\subfloat[]{
\includegraphics[scale=0.35]{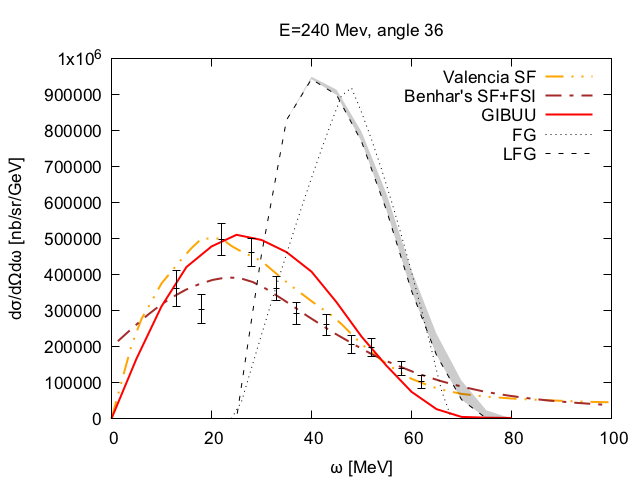}
\label{fig:30-50_ac}
}
\subfloat[]{
\includegraphics[scale=0.35]{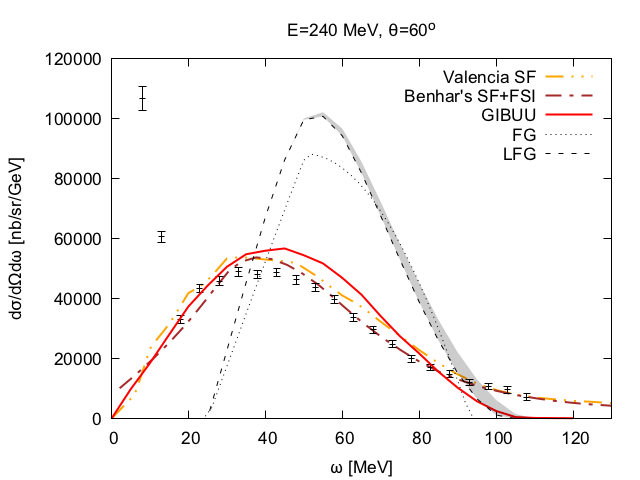}
\label{fig:30-50_ad}
}
\hspace{0mm}
\subfloat[]{
\includegraphics[scale=0.35]{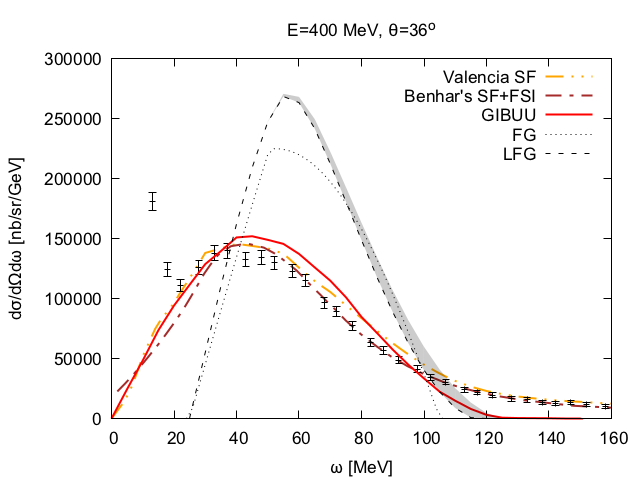}
\label{fig:30-50_ae}
}
\caption{Comparison of electron scattering models in Region I. For the local Fermi gas the relativistic kinematics is employed and the difference from nonrelativistic version is shown as a grey band.}
\label{fig:30-50_a}
\end{figure*}

In this region the longitudinal part of the cross section overwhelmingly dominates (see Table. \ref{table:electrons2}), and contributions for $2p2h$ and $\Delta$ mechanisms are supposed to be negligible, hardly influencing our analysis. Thus we can compare models to the data directly. On the other hand, it is a region sensitive to collective effects, which are clearly visible in the data (Fig. \ref{fig:30-50_aa}-\ref{fig:30-50_ae}). With so small energy transfers, we are at the verge of usability of the models and the data points lying on the left of the QE peak are usually not described correctly using just spectral functions. Moreover, as can be seen on Fig. \ref{fig:30-50_aa}, \ref{fig:30-50_ac} it is sometimes difficult to discern the QE peak from the giant resonances.

In these kinematical setups the importance of nuclear effects is clearly visible. LFG predicts the QE peak sometimes almost twice as high as the data and shifted (up to 15 MeV). GFG predictions are slightly lower but also clearly overestimates the experimental points. Benhar's SF misplaces the QE peak (even by 20-25 MeV) and overestimates it (up to $10\%$ effect), confirming that not only the hole state should be properly described inside the nucleus. When applying complete models: Valencia SF, Benhar's SF+FSI and GiBUU, the results are close to the data and all the models work remarkably well (Fig. \ref{fig:30-50_aa}-\ref{fig:30-50_ae}). 

Valencia SF and Benhar's SF+FSI performance is probably slightly better then GiBUU. GiBUU overestimates the QE peak and its right slope. Also we observe that Valencia SF and Benhar's SF+FSI produce a long tail which is not present for GiBUU, however this influences only high energy transfer region. The tail would appear also for GiBUU if the collisional broadening was taken into account (here we show the results of the standard calculation which neglects it). 

In this region the usage of nonrelativistic kinematics for Valencia SF hardly influences the results. This can be observed for the LFG, where the difference between relativistic and nonrelativistic kinematics is small (see the grey band on Fig. \ref{fig:30-50_aa}-\ref{fig:30-50_ae}).

\subsection{II Region (energy transfer 50-125MeV)}

This is the most important region since it corresponds to the highest $\frac{d \sigma}{dq d\omega}_{T2K}^{\text{osc}}$. There are plenty of data sets (18) to compare with, for various kinematical setups (both low and intermediate scattering angles).

First of all we notice that GiBUU, Benhar's SF+FSI and Valencia SF predictions stay in a very good agreement with the data in this region. For higher energy transfers (above the QE peak) Valencia predicts broader QE peak then the other two models, however this can be explained by nonrelativistic kinematics that is employed. GiBUU and Benhar's model in some cases underestimate the data (especially on the right side of the peak), however this can be partially explained by $2p2h$ contribution that is not included in the calculations, as will be discussed in Sec. \ref{sec:2p2h} (see Fig. \ref{fig:50-125ca},  \ref{fig:50-125ce} and \ref{fig:50-125cf}). The difference between models in the QE peak is rather small (few percent difference, GiBUU lying below). Lower energy transfer (below 50-75 MeV) is where the discrepancy is getting more pronounced. GiBUU in some cases slightly overesimates the data (e.g. Fig. \ref{fig:50-125cf}), while Benhar's SF+FSI and Valencia SF underestimate it (e.g. Fig. \ref{fig:50-125cc}). On the other hand, the reconstruction of cross section in the low energy region is beyond the scope of those models (giant resonances are well visible in the data). On Fig. \ref{fig:50-125be}-\ref{fig:50-125bf} one can see additional low-energy peaks whose structure might be described by collective modes.
 
Fermi gas models have a tendency to overestimate the QE peak. LFG predicts the QE peak far too high, for lower energy transfer almost twice high, going down to $40-60\%$ effect. It is also displaced for about $10-15$ MeV in the case of high-energy incoming electron and small scattering angle (see Fig. \ref{fig:50-125bd}, \ref{fig:50-125bf}, \ref{fig:50-125cb}). Changing the binding energy would move the QE peak position, however it is impossible to choose one value correct for all kinematics. 

If we neglect FSI, using Benhar's SF, the QE peak is shifted - up to 25 MeV - and slightly overestimated (what is especially pronounced for low scattering angle). However the latter effect is rather small (few percent). 

\begin{figure*}[h]
\centering
\subfloat[]{
\includegraphics[scale=0.35]{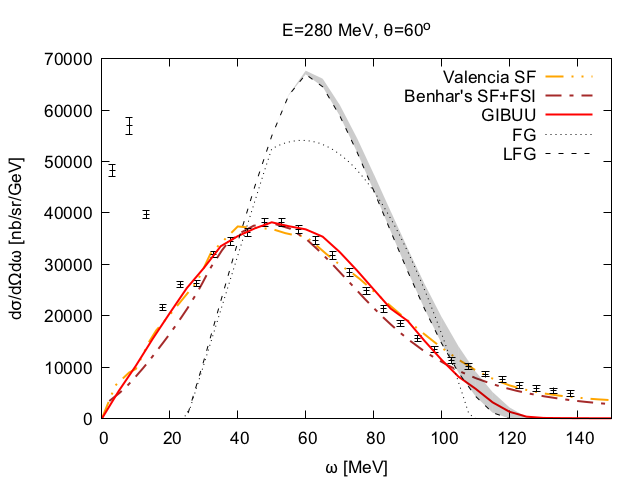}
\label{fig:50-125aa}
}
\subfloat[]{
\includegraphics[scale=0.35]{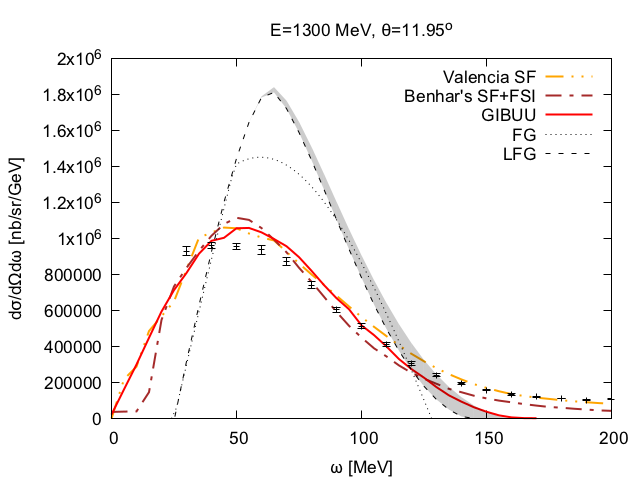}
\label{fig:50-125ab}
}
\hspace{0mm}
\subfloat[]{
\includegraphics[scale=0.35]{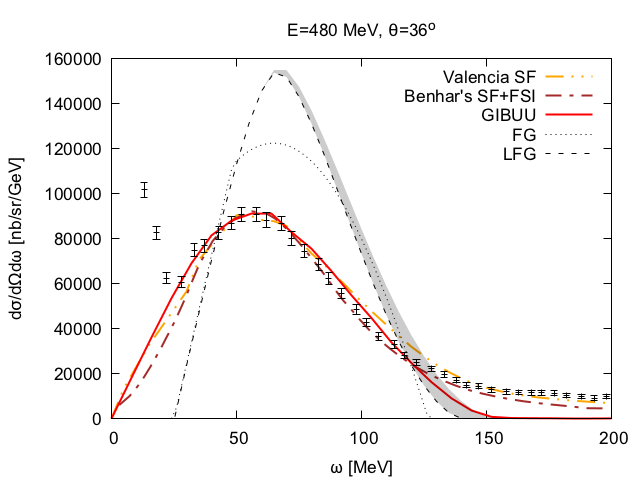}
\label{fig:50-125ac}
}
\subfloat[]{
\includegraphics[scale=0.35]{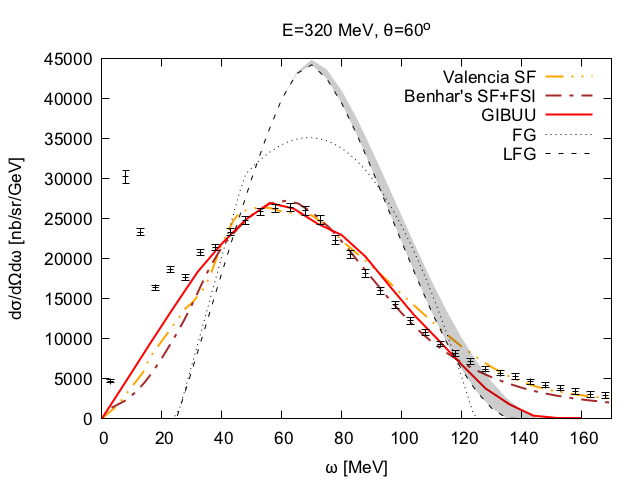}
\label{fig:50-125ad}
}
\hspace{0mm}
\subfloat[]{
\includegraphics[scale=0.35]{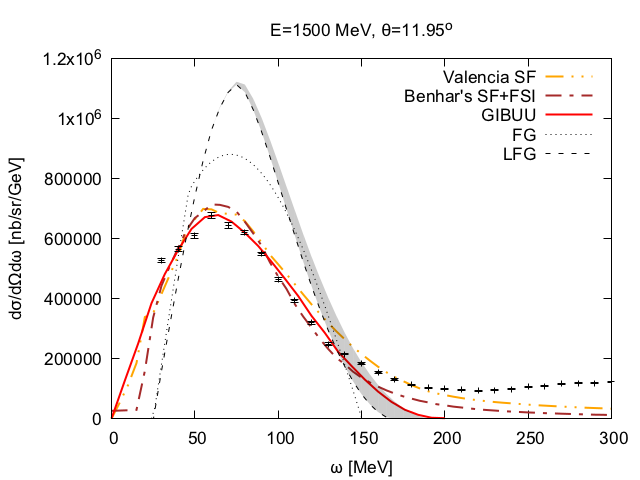}
\label{fig:50-125ae}
}
\subfloat[]{
\includegraphics[scale=0.35]{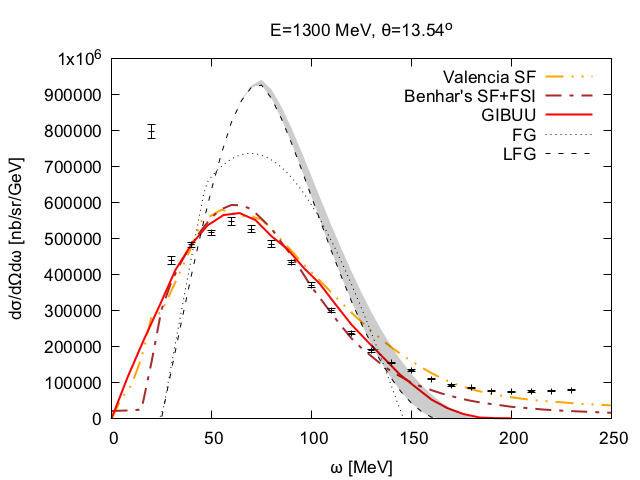}
\label{fig:50-125af}
}
\caption{Comparison of electron scattering models in Region II (for which the QE peak position is at $\omega \in (50,125)$ MeV).}
\label{fig:50-125a}
\end{figure*}

\begin{figure*}[h]
\centering
\hspace{0mm}
\subfloat[]{
\includegraphics[scale=0.35]{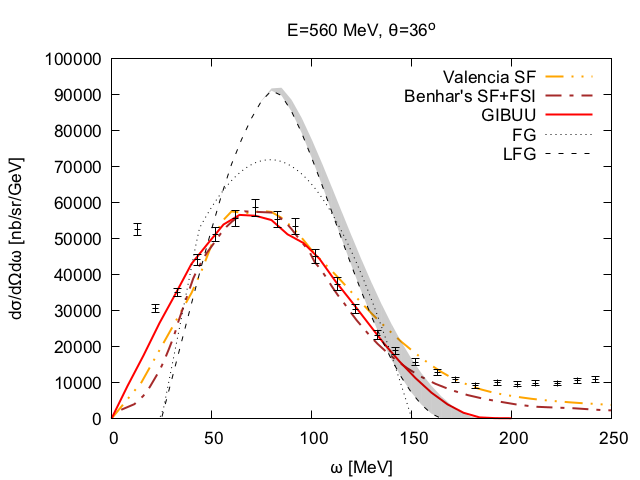}
\label{fig:50-125ba}
}
\subfloat[]{
\includegraphics[scale=0.35]{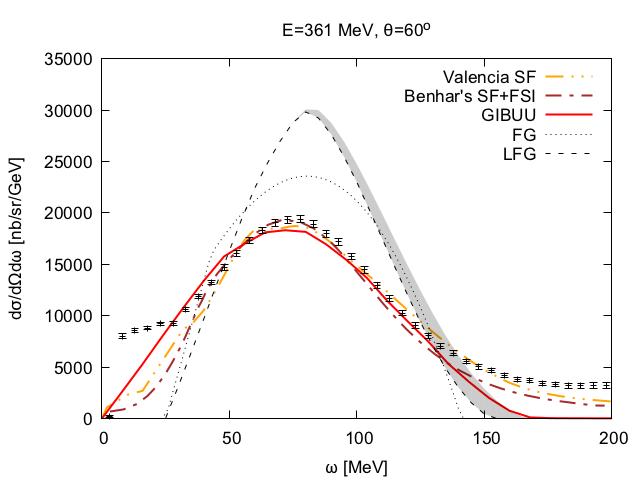}
\label{fig:50-125bb}
}
\hspace{0mm}
\subfloat[]{
\includegraphics[scale=0.35]{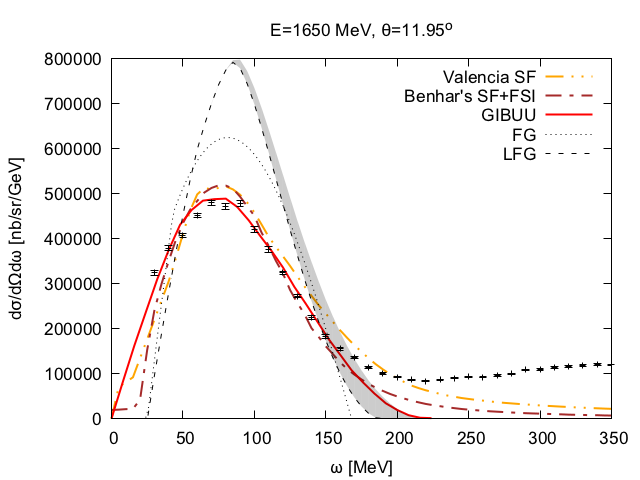}
\label{fig:50-125bc}
}
\subfloat[]{
\includegraphics[scale=0.35]{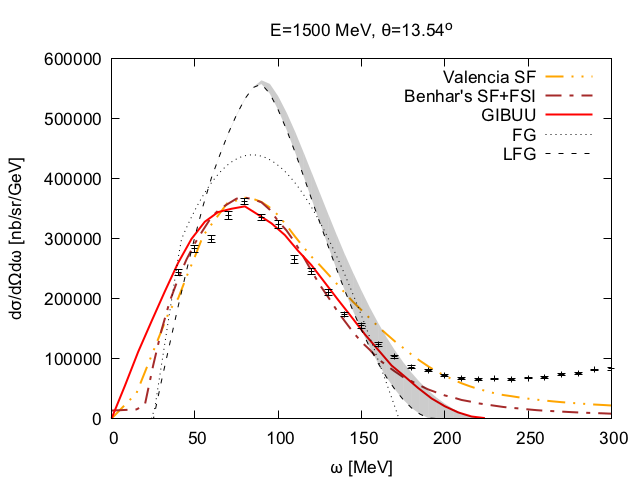}
\label{fig:50-125bd}
}
\hspace{0mm}
\subfloat[]{
\includegraphics[scale=0.35]{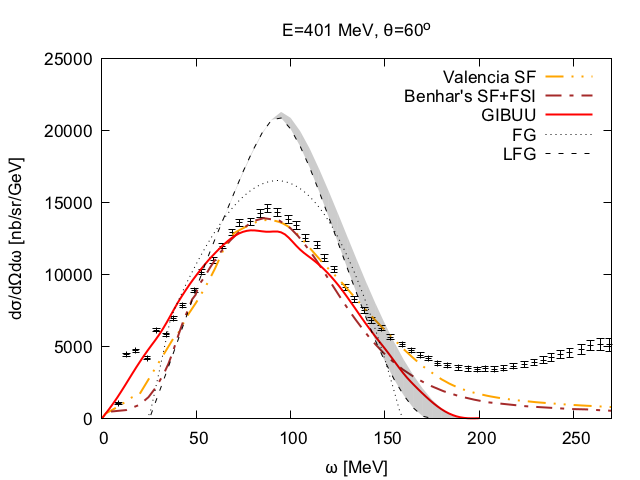}
\label{fig:50-125be}
}
\subfloat[]{
\includegraphics[scale=0.35]{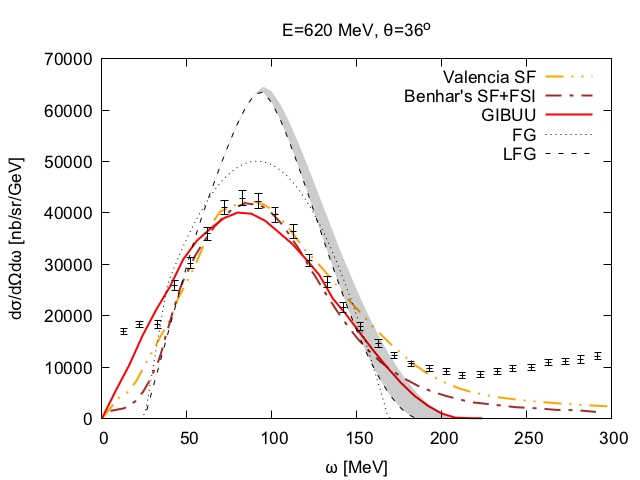}
\label{fig:50-125bf}
}
\caption{The same as Fig. \ref{fig:50-125a}.}

\label{fig:50-125b}
\end{figure*}

\begin{figure*}[h]
\centering
\hspace{0mm}
\subfloat[]{
\includegraphics[scale=0.35]{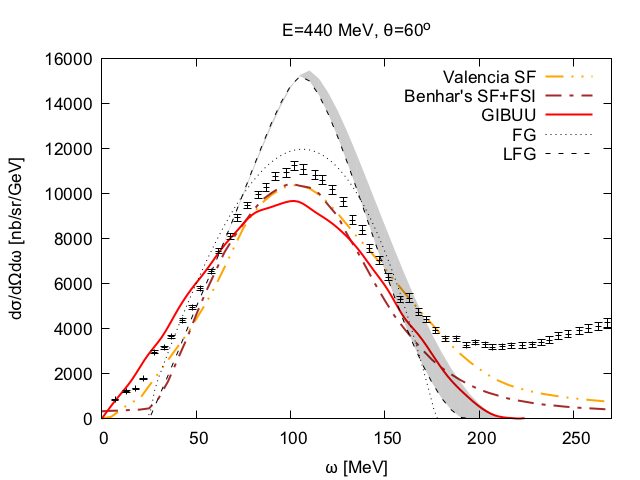}
\label{fig:50-125ca}
}
\subfloat[]{
\includegraphics[scale=0.35]{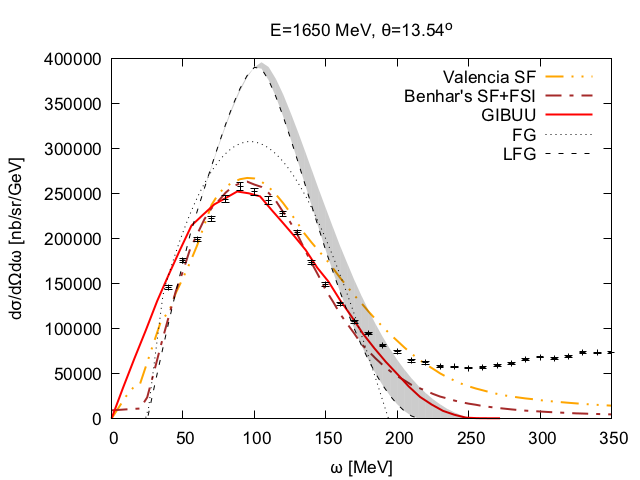}
\label{fig:50-125cb}
}
\hspace{0mm}
\subfloat[]{
\includegraphics[scale=0.35]{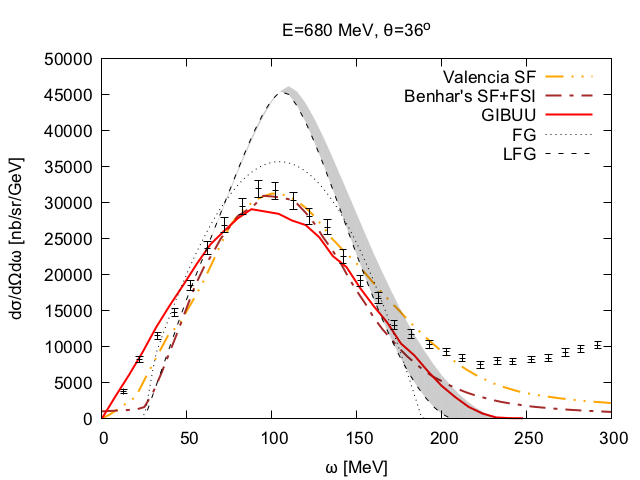}
\label{fig:50-125cc}
}
\subfloat[15]{
\includegraphics[scale=0.35]{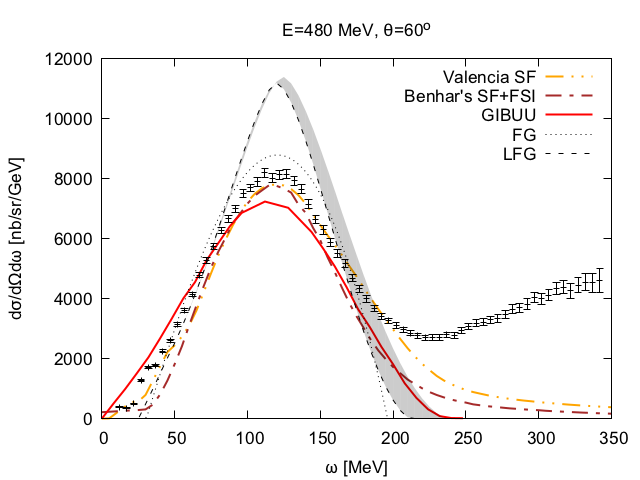}
\label{fig:50-125cd}
}
\hspace{0mm}
\subfloat[]{
\includegraphics[scale=0.35]{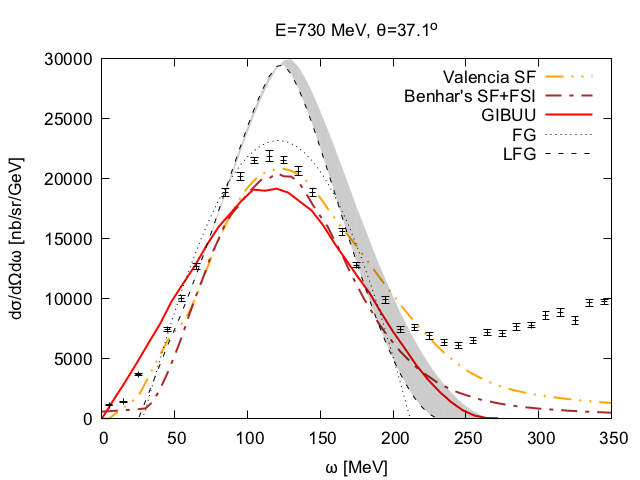}
\label{fig:50-125ce}
}
\subfloat[]{
\includegraphics[scale=0.35]{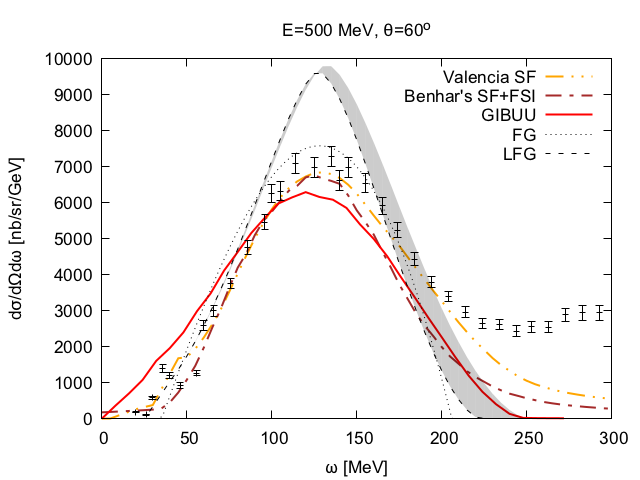}
\label{fig:50-125cf}
}
\caption{The same as Fig. \ref{fig:50-125a} and \ref{fig:50-125b}.}
\label{fig:50-125c}
\end{figure*}

As in Region I, Behnar's SF+FSI and Valencia SF predict a long tail spreading to the high energy transfer. They remarkably well recovers the data, as can be observed e.g. on Fig. \ref{fig:50-125a}. 

In this Region in general $\frac{\sigma_T}{\sigma_L}<1$ but as can be seen on Fig. \ref{fig:50-125cd}-\ref{fig:50-125cf} for $\frac{\sigma_T}{\sigma_L}\approx1$ other mechanisms start to be visible in the data as an enhancement in the cross section for large $\omega$.

\subsection{Region III (energy transfer 125-200 MeV)}
In this region $\sigma_T \sim \sigma_L$ and $2p2h$ and $\Delta$ mechanisms are expected to overlap with the QE peak in a significant way. However, as will be shown in Sec. \ref{sec:2p2h}, the effect is not drastic in the vicinity of the peak. It influences mostly the right slope of the QE peak, hardly changing the height of the peak itself.

The higher energy transfer is, the less visible nuclear effects become. Comparing to Region I and II, the differences between the Fermi gas and other models are less pronounced. Surprisingly, GFG gives predictions very similar to models which contain nuclear corrections. As we have seen in Region I and II, LFG overestimates the peak. Here the effect is not so drastic, yet clearly visible, about $35\%$ of discrepancy.

Benhar's SF also slightly misplaces the peak (up to 15 MeV towards high energies), however it does not overestimate it. The inclusion of FSI, which caused quite a strong quenching in Region I and II, here produces rather a small effect. The peak gets shifted towards lower energies because of the real part of the optical potential $U_V$.

\begin{figure*}[h]
\centering
\subfloat[]{
\includegraphics[scale=0.35]{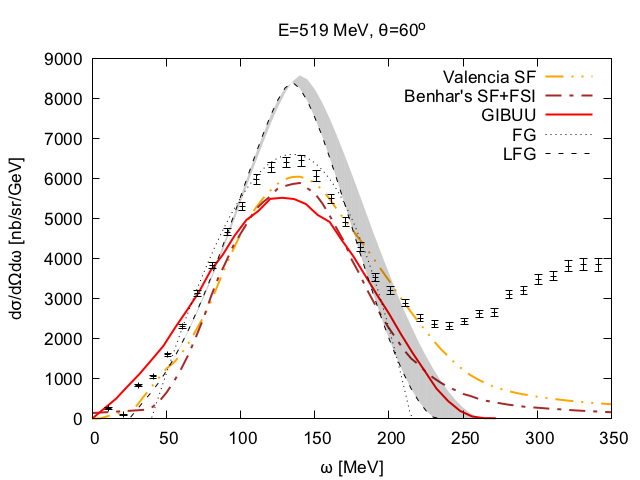}
\label{fig:125-200_aa}
}
\subfloat[]{
\includegraphics[scale=0.35]{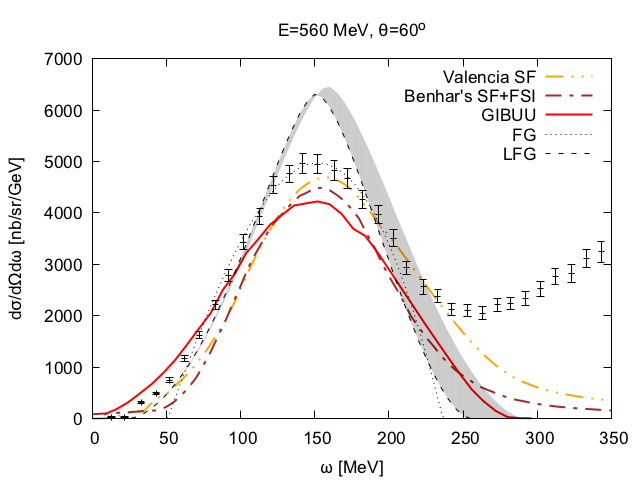}
\label{fig:125-200_ab}
}
\hspace{0mm}
\subfloat[]{
\includegraphics[scale=0.35]{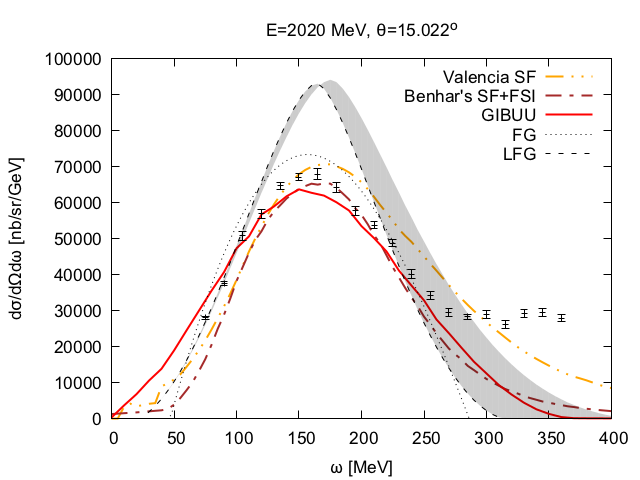}
\label{fig:125-200_ac}
}
\subfloat[]{
\includegraphics[scale=0.35]{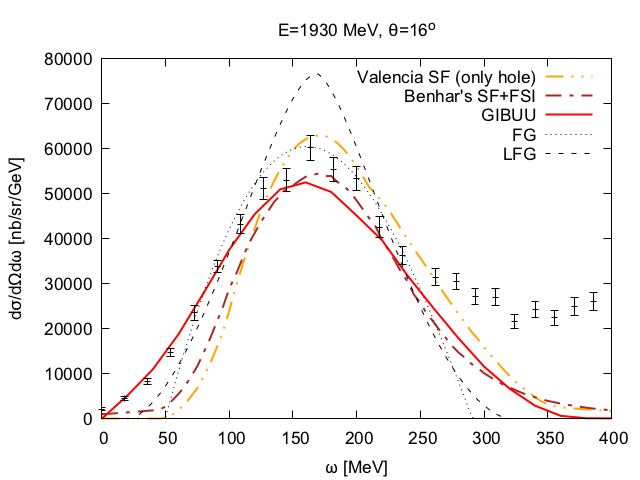}
\label{fig:125-200_ad}
}
\hspace{0mm}
\subfloat[]{
\includegraphics[scale=0.35]{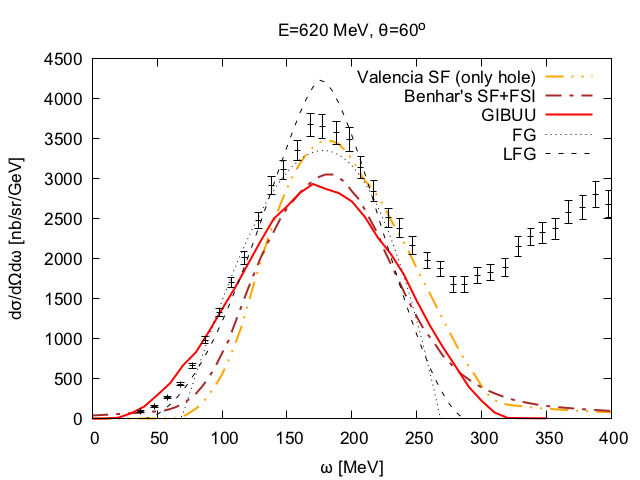}
\label{fig:125-200_ae}
}
\subfloat[]{
\includegraphics[scale=0.35]{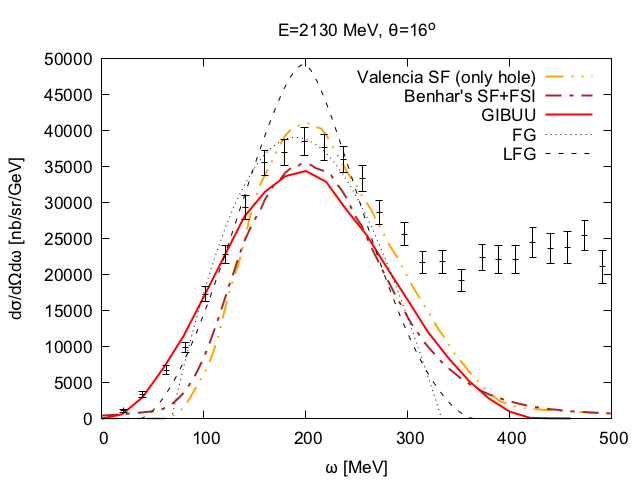}
\label{fig:125-200_af}
}
\caption{Comparison of electron scattering models in Region III (for which the QE peak position is at $\omega \in (125,200)$ MeV).}
\label{fig:125-200_a}
\end{figure*}

\begin{figure*}[h]
\centering
\subfloat[]{
\includegraphics[scale=0.35]{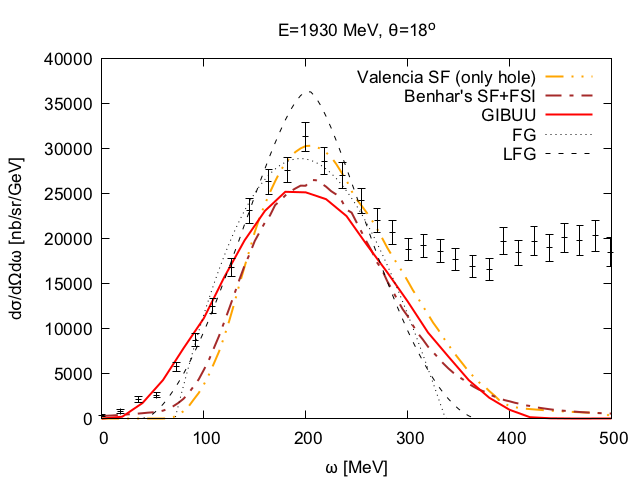}
\label{fig:125-200_ba}
}
\subfloat[]{
\includegraphics[scale=0.35]{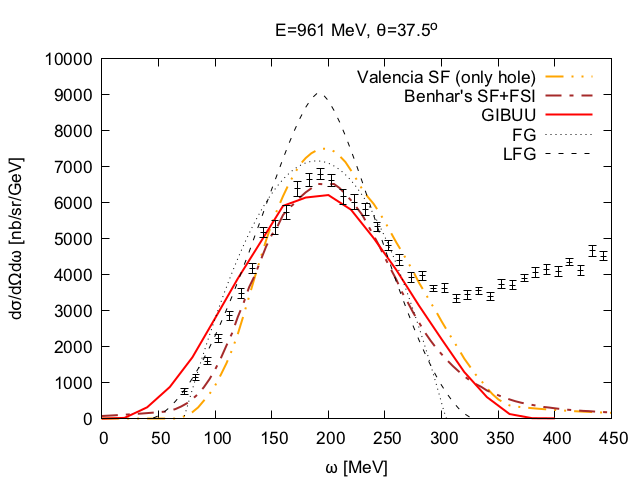}
\label{fig:125-200_bb}
}
\hspace{0mm}
\subfloat[]{
\includegraphics[scale=0.35]{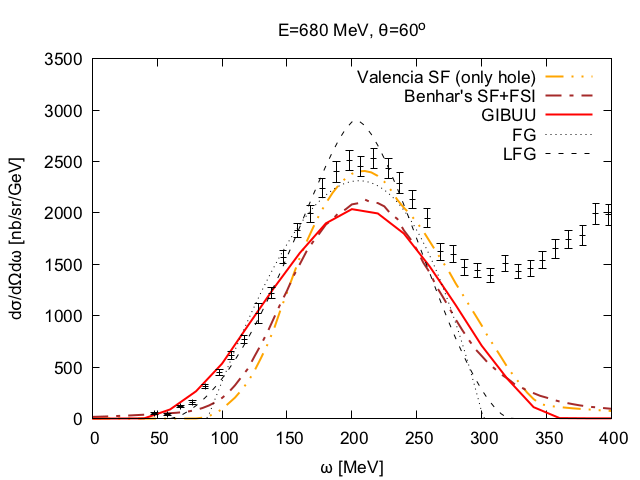}
\label{fig:125-200_bc}
}
\caption{The same as Fig. \ref{fig:125-200_a} and }
\label{fig:125-200_b}
\end{figure*}

In this region, because of large relativistic effects, for all the cases but three with lowest energy transfer (Fig. \ref{fig:125-200_aa}, \ref{fig:125-200_ab}, \ref{fig:125-200_ac}) we use an approximated version of Valencia SF, where we neglect the particle spectral function (or in other words FSI). The results are shown on Fig. \ref{fig:125-200_ad}, \ref{fig:125-200_ae}, \ref{fig:125-200_af}, \ref{fig:125-200_b}. Comparing to GiBUU and Benhar's SF+FSI model, the peak is around 10-15\% higher and slightly broader on the right side of the QE peak, which makes it closest to the data, however not leaving space for the inclusion of $2p2h$ and $\Delta$ effects. 

GiBUU and Benhar's SF+FSI models give very similar predictions of the height of the QE peak (a few percent difference GiBUU lying lower), but different position. We observe that GiBUU peak is much broader, and for this reason in general for lower energy transfer better fits with data (sometimes overestimating it). Benhar's SF+FSI and Valencia SF underestimate data on the left side of the QE and this cannot be explained by the $2p2h$ mechanism. This difference between the models was already visible in Region II.

\section{Discussion}\label{sec:discussion}
One encounters various practical problems when trying to perform a more quantitative comparison of the models. Recovery of the QE peak means not only its position, height and width. The peak's shape also changes, from a fully symmetric to asymmetric (e.g. Fig. \ref{fig:30-50_ad}, \ref{fig:50-125ba}). Moreover, we do not have enough ''direct information'' even to measure those basic nuclear quantities. For the data from Region I and partially from Region II we cannot calculate the peak's width because experimental points for low-energy transfer are obscured by the existence of giant resonances. On the other hand, for the data from Regions II and III the contributions from $2p2h$ and $\Delta$ broaden the peak, changing also its height and occulting the shape.

In order to overcome those obstacles, we attempt to subtract $2p2h$ and $\Delta$ contributions from the data, using a model from \cite{Megias:2016lke}. We are well aware of the fact that this is only an approximated procedure, because a prediction for $2p2h$ and $\Delta$ mechanisms is model dependant and there may be non-neglible interefence effects. However, we hope that at least basic properties of the models can be analyzed this way.

We will concentrate on two characteristics of the QE peak: its position and height. Subtracting $2p2h$ and $\Delta$ contributions affects mainly the width of the QE peak, changing the QE peak's height only for the high energy transfer by less then 10\% percent (according to the model from \cite{Megias:2016lke}).

The last point raised in the discussion, in Sec. \ref{sec:comp_lfg}, will be the comparison of the QE peak's height and width (defined as the peak's width in the middle of the height) predicted by various models and LFG. 

\subsection{2p2h and $\Delta$}\label{sec:2p2h}
$2p2h$ mechanism, also called meson-exchange current (MEC), describes a situation when a lepton interacts with two nucleons from the ground state, creating another pair of nucleons. In some kinematical regions $2p2h$ and $\Delta$ mechanisms may give noticeable contribution to the QE peak. These new physical mechanisms overlap with the QE peak, as can be observed with the bare eye when looking at the data: the cross section grows instead of diminishing for higher energy transfer (above the QE peak), see e.g. Fig. \ref{fig:125-200_a}. From the theoretical perspective, this large contribution is connected to the fact that the transversal response function is giving sizable contribution to the cross section.

In Tables \ref{table:electrons1} and \ref{table:electrons2} we show a ratio $\frac{\sigma_T}{\sigma_L}$ in the QE peak for every data set. In general, for low energy transfer the ratio is small ($< 1$), increasing for higher energies. A low ratio means that $R_T$ is suppressed and thus we suspect that $2p2h$ and $\Delta$ contributions will be relatively small (or even negligible). This refers mainly to the data from Region I and partially from Region II. However, for high $\frac{\sigma_T}{\sigma_L}$ ratio no such conclusion can be drawn. In some cases (however not necessarily) we might expect larger effect.

We must also note that Benhar's spectral function includes initial nucleons' correlations which are partially responsible for $2p2h$ contribution so we should expect the sum of the QE and $2p2h$ mechanisms to exceed experimental data. The same happens with Valencia spectral function which also contains some $2p2h$ dynamics.

\begin{figure*}[h]
\centering
\subfloat[]{
\includegraphics[scale=0.35]{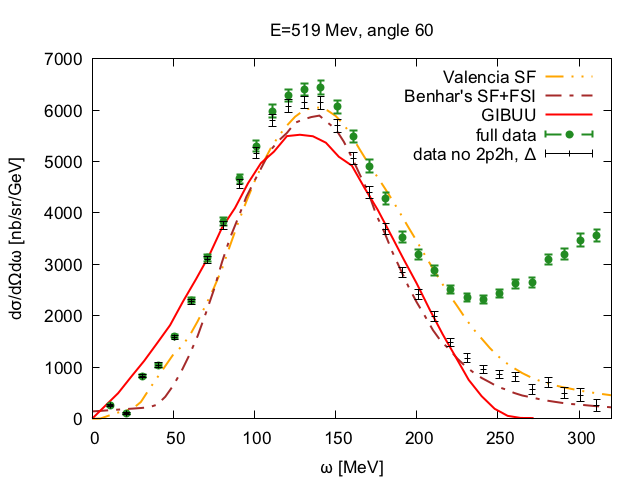}
\label{fig:2p2h_1}
}
\subfloat[]{
\includegraphics[scale=0.35]{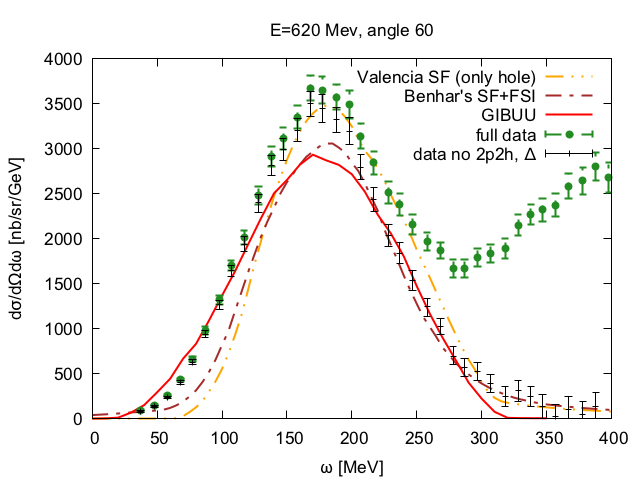}
\label{fig:2p2h_2}
}
\caption{A comparison of three models - GiBUU, Valencia SF and Benhar's SF+FSI, with the data. We subtract $2p2h$ and $\Delta$ contributions using results from \cite{Megias:2016lke} from the data points. Both data sets (before and after subtraction) are shown. Left panel: full Valencia SF is used, right panel: approximated Valencia SF is used (without FSI).}
\label{fig:2p2h}
\end{figure*}

To see how large the effect may be, on Fig. \ref{fig:2p2h_1}, \ref{fig:2p2h_2} we show two examples (from Region III), where the $2p2h$ and $\Delta$ play an important role. We may see that the models (mainly Benhar's SF+FSI and GiBUU)  underestimate the data, especially on the right side of the QE peak. After approximate subtraction of $2p2h$ and $\Delta$ contributions from the data, the agreement becomes much better. Still, however, in these particular situations the height of the QE peak is underestimated for GiBUU: $2p2h$ and $\Delta$ have much smaller effect in the peak's vicinity. Valencia SF overestimates the right side of the QE peak in both cases but from different reasons.  On Fig. \ref{fig:2p2h_1} it is caused by nonrelativistic kinematics while on Fig. \ref{fig:2p2h_2} it is because approximated Valencia SF is used.

\subsection{QE peak's height and position}
\begin{figure*}[h]
\centering
\includegraphics[scale=0.35]{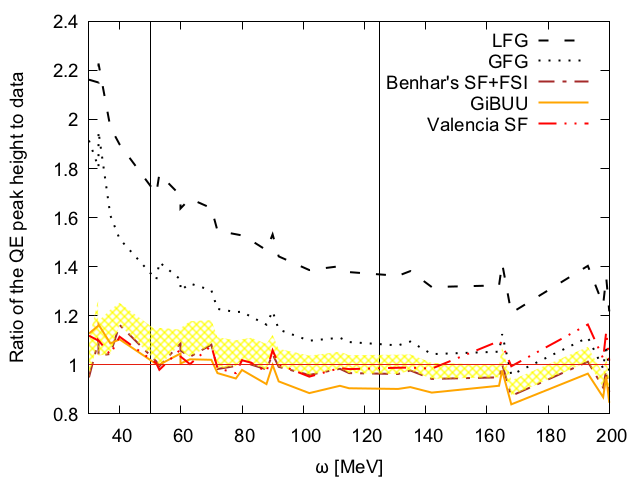}
\caption{Ratio $\frac{d\sigma}{d\Omega d\omega}^{model} /  \frac{d\sigma}{d\Omega d\omega}^{data}$ in the QE peak for various models. A band above Benhar's SF+FSI line shows the difference from Benhar's SF alone. Two vertical lines at 50 and 125 MeV tentatively mark Region I, II, III.}
\label{fig:heightDATA}
\end{figure*}

On Fig. \ref{fig:heightDATA} we show a ratio $\frac{d\sigma}{d\Omega d\omega}^{model} /  \frac{d\sigma}{d\Omega d\omega}^{data}$ of the QE peak height for data and various models, as a function of the QE peak position (in energy transfer). 

We see how it changes drastically for both Fermi gas models. They overestimate the QE peak height very strongly especially for low energy transfer. The GFG tends to quite good agreement with the data for higher energies, however the LFG still overestimates it by $30-40\%$ for $\omega>125$ MeV.

For other models the ratio does not depend so strongly on the values of the energy transfer and their predictions are close to each other. GiBUU slightly overestimate the data for low energies and then underestimates it for moderate and high energies. This behaviour can be partially (although not fully) explained by an additional $2p2h$ contribution. Valencia SF gives almost the same results as Benhar's SF+FSI, although for high energy transfer - where the approximation of neglecting the particle SF is used - it overestimates the data.

\begin{figure*}[h]
\centering
\subfloat[Local Fermi gas]{
\includegraphics[scale=0.35]{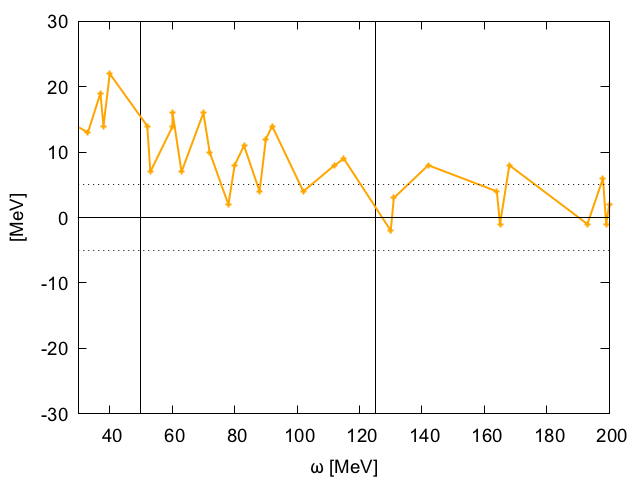}
\label{fig:peakLFG}
}
\subfloat[Global Fermi gas]{
\includegraphics[scale=0.35]{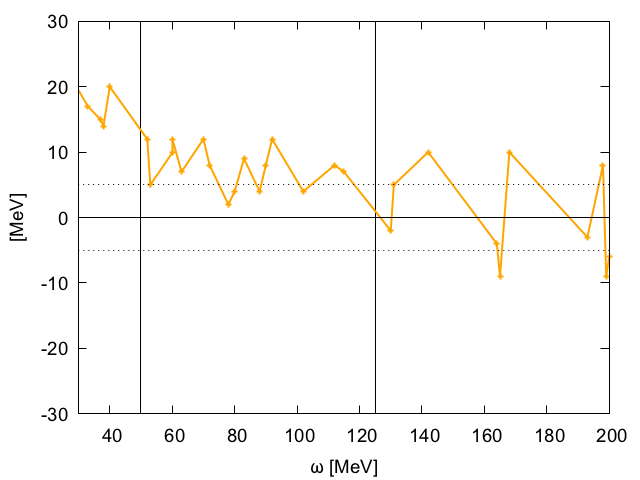}
\label{fig:peakFG}
}
\hspace{0mm}
\subfloat[SF+FSI Benhar]{
\includegraphics[scale=0.35]{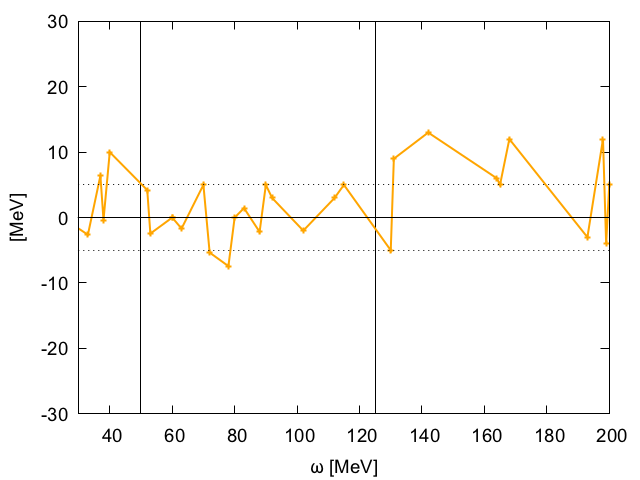}
\label{fig:peakBenhar}
}
\subfloat[GiBUU]{
\includegraphics[scale=0.35]{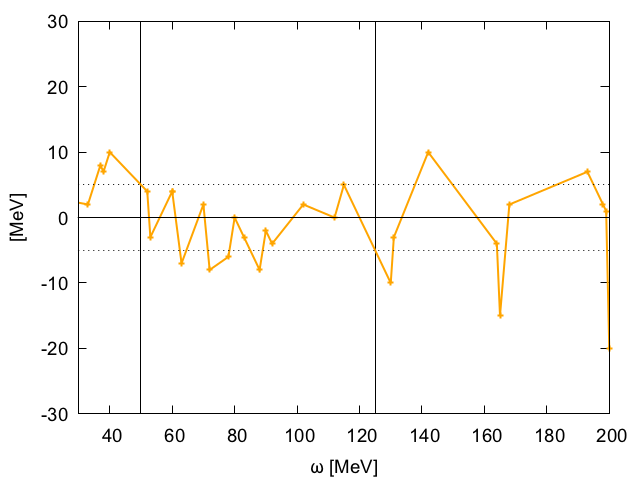}
\label{fig:peakGibuu}
}
\hspace{0mm}
\subfloat[SF Valencia]{
\includegraphics[scale=0.35]{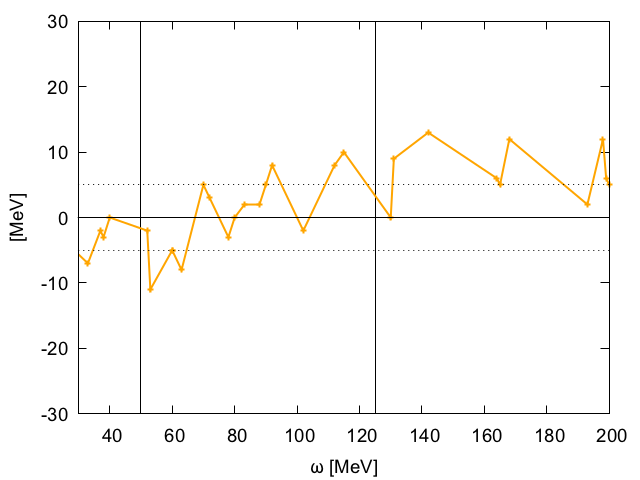}
\label{fig:peakValencia}
}
\caption{Position of the QE peak with respect to the data. Because of the discrete nature of experimental points, an error band of 5 MeV is also shown with dotted line. Two vertical lines at 50 and 125 MeV tentatively mark Region I, II, III.}
\label{fig:peakPosition}
\end{figure*}

On Fig. \ref{fig:peakPosition} we plot the position of the QE peak according to various models with respect to the data as a function of the energy transfer. In Region I Fermi gas models shift the peak by more then 10-20 MeV towards higher energy transfer. In Region II the effect is smaller and already comparable with other models predictions. Surprisigly, for Region III the LFG seems to be the closest to the data. We must however remember that here $2p2h$ might be responsible for moving slightly the peak towards higher energy transfer.

\subsection{Comparison to LFG for the QE peak width and height}\label{sec:comp_lfg}
\begin{figure*}[h]
\centering
\subfloat[Height]{
\includegraphics[scale=0.35]{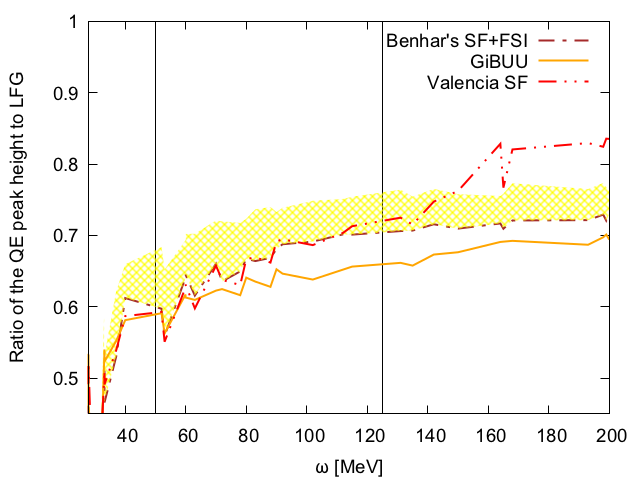}
\label{fig:heightLFG}
}
\subfloat[Width]{
\includegraphics[scale=0.35]{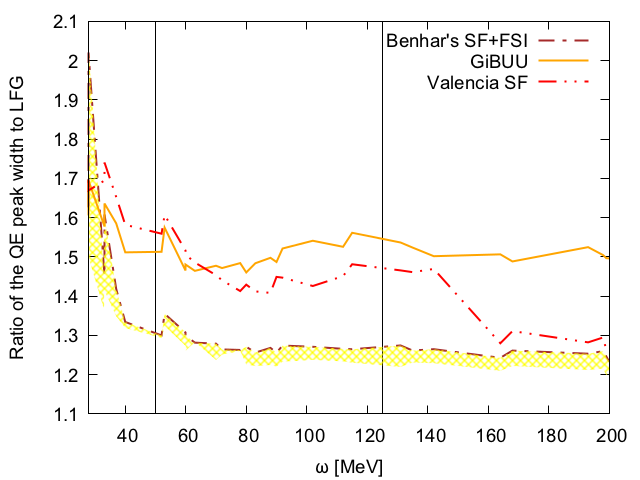}
\label{fig:widthLFG}
}
\caption{On the left: ratio $\frac{d\sigma}{d\Omega d\omega}^{model} /  \frac{d\sigma}{d\Omega d\omega}^{LFG}$ in the QE peak for models: Benhar's SF+FSI, GiBUU and Valencia SF. On the right: ratio $\frac{\omega_R^{model}-\omega_L^{model}}{\omega_R^{LFG}-\omega_L^{LFG}}$ where $\omega_R$, $\omega_L$ are the energy transfer values in the middle of the peak's height. A band above/below Benhar's SF+FSI line shows the difference from Benhar's SF alone. For $\omega>150$ MeV we use approximated Valencia SF model. Two vertical lines at 50 and 125 MeV tentatively mark Region I, II, III.}
\end{figure*}

As it was said before, a Fermi gas model is a reasonable benchmark to estimate the importance of nuclear effects. Hence we want to show how various nuclear corrections influence the height and width of the QE peak and how this depends on value of the energy transfer. Even though for comparison with the data we were not able to perform an analysis of the peak's width, here we can do it. On Fig. \ref{fig:heightLFG} we show an analogical plot to Fig. \ref{fig:heightDATA}, for the ratio $\frac{d\sigma}{d\Omega d\omega}^{model} /  \frac{d\sigma}{d\Omega d\omega}^{LFG}$. The difference between LFG and all the models is more pronounced at low energy transfer.  Benhar's SF+FSI causes stronger quenching of the QE peak with respect to Benhar's SF model (the difference is marked by a band). GiBUU predictions lie even lower, however the general behaviour is the same for those models: starting from ratio $\approx 0.5$ and going up to $\approx 0.7-0.8$ for 200 MeV. Valencia SF for low energy transfer (Region I) works in a very similar way to GiBUU and then is getting closer to Benhar's SF+FSI. In Region III above 150 MeV its preditions are higher than the other two models because we neglect the particle spectral function. Its introduction would cause quenching of the peak, a similar effect that can be seen in Benhar's model.

The behaviour of the QE peak's width is shown on Fig. \ref{fig:widthLFG}. We define it as $\omega_R - \omega_L$ where $\omega_{R,L}$ correspond to the energy transfer on the right/left of the QE peak, where the cross section is a half of that at the peak.  Again the largest nuclear effects are visible for Region I. There we can see that Benhar's SF+FSI, GiBUU and Valencia SF give very similar results. In Region II, however, Benhar's SF+FSI is narrower. Looking at the data sets, we see that the difference comes from the low-energy transfer region (below the QE peak).  For higher energy transfer we should remember that relativistic effects are already visible and Valencia SF peak's width is overestimated. We can also observe that when neglecting the particle spectral function, Valencia SF gives narrower peak (for energy transfer $\omega>150$ MeV).

The general energy dependance for all three models is similar (the ratio diminishes for growing energies comparing to Region I). The speading of the peak caused by the FSI is also well visible, and this effect grows slightly with the energy transfer.

\section{Conclusions}\label{sec:conlusions}
We have analyzed how various models of the electron-nucleus scattering behave in the kinematical region important for the T2K experiment. We have focused on the QE region, surveying models of nuclear corrections expressed in terms of spectral functions.

The conclusions drawn from the electron data analysis cannot be straightforwardly applied to the neutrino case. This is not only because of the different  interaction vertex (neutrino-nucleus interaction has an additional axial part), but mainly because of the fact that we deal with a flux of neutrinos. Integration over incoming neutrino beam blurs the picture. As we have seen, each electron-scattering model works differently for various kinematical regions (depending on the scattering angle and the energy) and some effects may be enhanced or cancelled out (e.g. an overestimation in one region and underestimation in another may accidentally sum up to a good result). We thus present general conclusions for every model, aiming to indicate those nuclear effects that should be important (and visible) for neutrinos.

Local Fermi gas overestimates the height of the QE peak in every region. The effect grows with the decreasing energy transfer, starting from a $20-40\%$ in Region III, through $40-80\%$ in Region II, up to $100\%$ in Region I. This effect cannot cancel out, however its intensity depends on the flux profile. GFG, comparing with the LFG, recovers data much better. It does not overestimate the QE peak so strongly - however the effect is also fairly visible. In both models the position of the peak is sometimes shifted (even by 25 MeV) but this is important only for low energy transfer and depends strongly on the scattering angle: one cannot choose the binding energy suitable for all the setups. The peak's shape is also far from the experimental one, lacking spread to higher energies. An inclusion of various nuclear corrections is needed to bring the predictions closer to reality. 

Benhar's SF causes quenching of the QE peak, still overestimating it (up to $10\%$ effect which is more pronounced for low energy transfer; for Region II it is only a few percent effect). It displaces the peak's position (up to $25$ MeV). However its shape is much closer to the experimental data, what can be especially observed in Region II where the spectral function recovers a high energy tail present in the data.

The shortcomings of Benhar's SF are healed by introducing FSI. The peak is shifted and slightly suppressed so that it matches the expermental points almost perfectly in the QE peak region. However, for Region II and III at the lower energy transfer (the left slope of the QE peak) the cross section is underestimated.

Valencia SF describes the data well, in a very similar way to Benhar's SF+FSI. Its main drawback is the nonrelativistic treatment of the kinematics, which forces us to neglect particle spectral function for higher energy transfer (Region III). This approximation leads to slightly overestimated QE peak.

Also GiBUU predictions recover the experimental data with a good precision. GiBUU slightly overestimates the QE peak height in Region I and underestimates it in Region II and III, however one cannot say exactly how much because of the unknown $2p2h$ contribution. Still it is below $10\%$ effect.  The low-energy part of the QE peak is a bit different with respect to Benhar and Valencia models: its broader width fits better with the data, sometimes, however, overestimating it (for Region II).

Among all the analyzed models, it is difficult to choose ''a winner'' which reproduces electron data with the greatest accuracy in all three regions. What is absolutely clear is that Fermi Gas models are not able to recover the data. There is a need of incorporating more sophisticated nuclear corrections. All formalisms presented in this paper have proved to work very well. From the computational point of view, apart from physical drawbacks and advantages, GiBUU and Benhar's SF+FSI need more CPU time than LFG, GFG or Valencia SF calculation.

There are good reasons to believe that when these models are used in neutrino event generators for a flux of T2K or similar one, they will introduce an error not exceeding a few percent. It is quite likely that this error smaller than the one coming from a limited knowledge of axial form factor of the weak nuclear current. It is also clear that a good model of $2p2h$ and $\Delta$ dynamics is necessary to model precisely the QE peak region.

\section*{Aknowledgements}
I am very grateful for the discussions with O. Benhar, U. Mosel and J. Nieves about models used in this comparison. Also I would like to thank A. Ankowski for providing me with essential files for the FSI implementation of Benhar's spectral function and many useful hints. And last but not least, I thank to J. T. Sobczyk for giving me the inspiration to write this paper and many comments on the way.
\bibliography{electrons}
\end{document}